%% file: draft_ver4.tex
\documentclass[11pt]{article}
\input{setting/setting.tex}

\usepackage{mathrsfs}
\usepackage{latexsym}
\usepackage{wrapfig}
\usepackage{tabularx}
\usepackage[normalem]{ulem}
\usepackage{feynmp-auto}
\usepackage{mathtools}
\usepackage{tikz}
\usetikzlibrary{decorations.pathreplacing}

\newcommand{\der}{\partial}
\newcommand{\no}{\nonumber}
\newcommand{\re}{\mathrm{Re}}
\newcommand{\im}{\mathrm{Im}}
\newcommand{\Mpl}{M_{\rm pl}}
\newcommand{\Lag}{\mathcal{L}}
\newcommand{\sys}{\mathcal{S}}

\newcommand{\cov}{\mathcal{M}}

\begin{document}

	\begin{titlepage}
		\setcounter{page}{1} \baselineskip=15.5pt 
		\thispagestyle{empty}

		\begin{center}
            \fontsize{18}{18}\textbf{False and genuine decoherence in the early universe:}\\
            \fontsize{16}{18}\textbf{a local observer and time-averaged observables}
        \end{center}		
		\vskip 15pt
		\begin{center}
			\noindent
			{\fontsize{12}{18} \selectfont 
                Fumiya Sano$^{a,b,}$\footnote{\href{mailto:sanof.cosmo@gmail.com}{sanof.cosmo@gmail.com}}, 
                Junsei Tokuda$^{c,}$\footnote{\href{mailto:junsei.tokuda@mcgill.ca}{junsei.tokuda@mcgill.ca}}}
		\end{center}
		
		\begin{center}
			\vskip 8pt
            $^a$ \textit{Department of Physics, Institute of Science Tokyo, Tokyo, 152-8551, Japan} \\
			$^b$ \textit{Center for Theoretical Physics of the Universe, Institute for Basic Science,\\Daejeon, 34126, Korea}\\
            $^c$ \textit{Department of Physics, McGill University, Montr\'{e}al, H3A 2T8, QC Canada}\\
            \end{center}
		
		\vspace{0.4cm}
            \begin{center}\textbf{Abstract}\end{center}
            We study quantum decoherence of curvature perturbations at superhorizon scales caused by the gravitational nonlinearities. We show that cubic gravitational couplings, constrained by the spatial diffeomorphism invariance, lead to infrared (IR) and ultraviolet (UV) divergences in the decoherence rate at one loop. These divergences arise respectively from deep IR fluctuations, which appear as background modes to a local observer, and from violent zero-point fluctuations in the deep UV. We argue that these divergences are unobservable, as they vanish when proper observables are considered. We consider correlators defined using the geodesic distance for IR divergences and time-averaged correlators for UV divergences. To incorporate these observer's perspective, we propose to use in effective quantum state, defined in terms of actual observables, as a more appropriate probe of the quantum coherence of the system measured by an observer. We then evaluate the finite decoherence rate induced by superhorizon environment during inflation and in the late universe.

		
	\end{titlepage} 
	
	
	\newpage
	\setcounter{tocdepth}{2}
	\setcounter{page}{2}
	{
		\tableofcontents
	}
	
	\newpage


\begin{fmffile}{diagram}

\setcounter{footnote}{0}


\section{Introduction}
Cosmic inflation~\cite{Starobinsky:1980te,Sato:1980yn,Guth:1980zm,Linde:1981mu} amplifies the quantum fluctuations of the metric at superhorizon scales, providing
quantum seeds for cosmological perturbations in the late time universe. 
Inflation may offer a useful tool for probing quantum nature of gravity, and
there has been growing interest in the loss of quantum behavior of inflationary fluctuations following pioneering works~\cite{Sakagami:1987mp,Brandenberger:1990bx,Grishchuk:1990bj,Brandenberger:1992jh,Albrecht:1992kf,Polarski:1995jg}.
{Various quantities have been discussed as a probe of quantum features of primordial fluctuations
such as entanglement entropy, quantum discord, purity,
complexity, and (temporal) Bell inequality, in addition to the reduced density matrix and Wigner function;
see, e.g.,~\cite{Campo:2005sv, Lim:2014uea,Maldacena:2015bha,Martin:2015qta,Martin:2016tbd,Choudhury:2016cso,Martin:2016nrr,Hollowood:2017bil,Kanno:2017dci,Bhattacharyya:2020rpy,Brahma:2020zpk,Martin:2021znx,Espinosa-Portales:2022yok,Martin:2022kph,Bhattacharyya:2024duw,Burgess:2024eng,Sou:2024tjv,Colas:2024ysu} for formulation of the measures, and~\cite{Burgess:2006jn,Campo:2008ju,Campo:2008ij,Nelson:2016kjm,Martin:2018zbe,Gong:2019yyz,Gong:2020gdb,Sou:2022nsd,Burgess:2022nwu,Ning:2023ybc,Colas:2024xjy,Takeda:2025cye,Lopez:2025arw} for evaluation of the quantum state for primordial metric perturbations.}
It is also argued in the context of quantum information theory that purity (or R\'enyi entropy as a generalization) characterizes maximal coherence of the state, which works as an upper bound for measures of quantum entanglement, discord, and coherence~\cite{Streltsov:2018}.

The purity is defined using a reduced density matrix $\rho_\text{red}$ for the system of interest. 
In this paper, we investigate the purity of fluctuations with wavelengths that are observed (or could be observed in future experiments) by a local observer. In this case, metric fluctuations at different scales act as the environment, involving both deep ultraviolet (UV) modes and infrared (IR) modes. The basic question we address concerns the subtleties of using $\rho_{\text{red}}$ as a diagnostic tool for the remaining quantum coherence of the system, especially when there exists a huge hierarchy between the scales of the system of interest and the environment.
This question was first raised in~\cite{unruh:2011fal}. Here, however, we examine a more general field theoretic framework, focusing in particular on cosmological settings.

In fact, it is known that one-loop calculations of purity or momentum-space entanglement entropy for a certain momentum scale sometimes suffer from infrared (IR) and/or ultraviolet (UV) divergences. 
{These divergences have been discussed in cosmological setup in, e.g.,~\cite{Nelson:2016kjm,Martin:2018zbe,Gong:2019yyz,Gong:2020gdb,Sou:2022nsd,Burgess:2022nwu,Lopez:2025arw} where a specific type of interactions was considered.} UV divergence exists in more generic setup, such as $g\phi^3$ theory in Minkowski spacetime with spacetime dimensions $D\geq 5$~\cite{Balasubramanian:2011wt}. Crucially, it is not canceled with the tree contributions from counterterms~\cite{Balasubramanian:2011wt} (see \cite{Burgess:2024heo} for more recent discussions).

These divergences imply that the purity defined by using $\rho_\text{red}$ could be sensitive to physics at very different scales. 
Hence, if this purity were a correct probe of the remaining quantum coherence of the system, no coherence would remain in the system because of the divergent entanglement with the environment at different scales, unless there were cancellation among contributions at different loop orders. This issue is closely related to recent discussions on the decoupling of UV physics~\cite{Burgess:2024heo}.

There are some expectations that these divergences may be absent from the beginning in properly defined correlators that would be measured by an actual local observer. 
For the IR side, the contributions from deep IR modes are expected to be merely gauge artifacts and thus cancel out when considering correlators measured by a local observer~\cite{Urakawa:2009my,Urakawa:2010kr,Tanaka:2011aj,Tanaka:2012wi,Tanaka:2013xe,Pajer:2013ana,Tanaka:2014ina,Tanaka:2017nff} (see also \cite{Tanaka:2013caa} for a review and references therein); a key ingredient is the use of geometric quantities, such as geodesic distance, to define locally measurable observables. {Moreover, it is suggested in \cite{Nelson:2016kjm} that decoherence induced by shorter wavelength modes might also be a gauge artifact.}
This treatment requires taking into account all the relevant interaction vertices that are necessary for recovering Maldacena's consistency relation~\cite{Maldacena:2002vr} (and its generalizations to correlators involving the conjugate momenta); however, to our knowledge, this complete treatment has not yet been achieved in the context of decoherence.
For the UV side, the divergence in the purity originates from the UV divergences in the equal-time limit of the Wightman functions in momentum space, which appear only in that limit. This point was suggested by previous works~\cite{Agon:2014uxa,Agon:2017oia,Burgess:2024heo,Bucciotti:2024lvb} and will be demonstrated explicitly in Section~\ref{sec:UVdiv}. Such divergences are absent in time-averaged correlators which reflect the finiteness of the time resolution of actual measurements.

The above discussions suggest the importance of considering the actual observables and accounting for all the relevant vertices in the computation of genuine quantum coherence of the system. However, it is unclear how to implement the first point into the computation of quantum state such as the reduced density matrix $\rho_\text{red}$. Even though the second point is merely a technical problem, it is desirable if there is an easy way to compute contributions from all the relevant interaction vertices.

In this paper, we study the quantum decoherence of the curvature perturbations at a certain superhorizon scale $\pm \mf{q}$ in the canonical single field inflation model. In this model, the sources of decoherence are gravitational nonlinear interactions which always exist. 
{We mainly consider the dynamics during inflation; however, we will also consider the late universe in Sections~\ref{sec:UVdiv} and \ref{sec:finite_deco}}.

To begin with, we overcome the second point by using the known facts that the purity at one-loop level is given by the integral of the product of tree-level cubic wavefunction coefficient $\psi_3$~\cite{Nelson:2016kjm} and that $\psi_3$ in the soft limit is fixed by the two-point functions of hard modes as a consequence of spatial diffeomorphism invariance~\cite{Pimentel:2013gza}. This considerably simplifies the one-loop computation of purity.  
We compute the purity using this method and confirm that the purity suffers from IR and UV divergences. It is then practically crucial to develop a method to compute the purity for a local observer who will not see such divergences.

We propose to consider {\it an effective quantum state} defined in terms of actual observables as a more appropriate probe of quantum coherence for the system measured by a local observer. 
A fundamental object we consider is the generating functional of $n$-point correlators measured by a local observer. From that functional, we define an effective reduced density matrix (or equivalently, an effective Wigner function), in terms of which an effective purity for a local observer is defined.

Applying our effective state formalism to the present setup, we yield the manifestly IR and UV finite effective purity. For the IR side, we adopt the correlators proposed in~\cite{Urakawa:2009my,Urakawa:2010kr,Tanaka:2011aj,Tanaka:2012wi,Tanaka:2013xe,Pajer:2013ana,Tanaka:2013caa,Tanaka:2014ina,Tanaka:2017nff} as genuine local observables. We highlight that the consistency relation plays an important role for the cancellation of IR divergences in the computation of effective purity.
For the UV side, we consider the correlators of operators smeared over time as observables. We demonstrate that the time smearing provides a robust mechanism for eliminating the spurious UV-divergent quantum decoherence of the system. 
Our result is also relevant for the purity lost due to the heavy physics. 
An important lesson we learn is that {\it the large entanglement predicted by the reduced density matrix does not necessarily imply a loss of coherence of the system.} 

We also find that the excited environment, such as the modes which is shorter than the system mode and cross the horizon mode during inflation, would serve as a genuine source of decoherence which is robust against the time averaging. We then evaluate finite decoherence rate caused by the superhorizon environment up to when the system modes about to re-enter the horizon during the radiation-dominated era. 
Note that our main results will be relevant for more generic setup including the study of primordial gravitational waves or the divergences of decoherence in flat space QFT; for instance, discussions of the wavefunction coefficient for a local observer are extended to the tensor modes in Appendix~\ref{sec:loe_tensor}.

This paper is organized as follows. In Section~\ref{sec:basic}, we explain our setup and introduce several basic quantities. We then utilize the consistency relation for wavefunction to compute the purity and find the IR and UV divergences. In Section~\ref{sec:eff_state}, we introduce the notion of effective quantum state for a local observer. 
In Section~\ref{sec:IRdiv}, we compute the effective purity for a local observer defined using the correlators proposed in~\cite{Urakawa:2009my,Urakawa:2010kr,Tanaka:2011aj,Tanaka:2012wi,Tanaka:2013xe,Pajer:2013ana,Tanaka:2013caa,Tanaka:2014ina,Tanaka:2017nff} and 
demonstrate the cancellation of IR diverfences. We highlight an important role of consistency relation for this cancellation. 
In Section~\ref{sec:UVdiv}, we compute an effective purity defined using the time-smeared observable and demonstrate how the UV divergence cancel out. We emphasize that there is no need to introduce an artificial cutoff in our computations.
Section~\ref{sec:finite_deco} is devoted to the evaluation of decoherence rate during and after inflation based on consistency relations. 
We conclude in Section~\ref{conclusion}. Several technical details are collected in appendices.

\section{Decoherence rate from consistency relations}\label{sec:basic}

\subsection{Reduced density matrix from wavefunction}\label{sec:reduced}
We consider the single field inflation in the comoving gauge, where only metric perturbations $\zeta, \gamma_{ij}$ are dynamical quantum variables. Here, $\zeta$ and $\gamma$ are comoving curvature perturbations and tensor perturbations, respectively. In the main text, we focus on $\zeta$ for simplicity. 
We suppose that the quantum state of curvature perturbations at the conformal time $\tau$ is a pure state $\ket{\Psi(\tau)}$.\footnote{Note that we will assume the initial state to be the adiabatic vacuum in the past infinity limit, though the results obtained in Section~\ref{sec:reduced} do not depend on this specific choice of initial state. In multi field models, additional degrees of freedom other than inflaton would serve as additional source of decoherence.}
We mainly consider the dynamics during inflation; however, we will also consider the late universe in Sections~\ref{sec:UVdiv} and \ref{sec:finite_deco}.

We are usually interested in correlation functions of $\zeta_{\mf{k}}$ with modes $\mf{k}$ in a specific domain $\sys$,\footnote{Even when specifying observable metric perturbations, their conjugate momenta are not unique due to freedoms of canonical transformation. 
Among such possibilities, the measurable conjugate momenta are limited to those which do not involve modes $\mf{k}$ outside the domain $\mathcal{S}$.}
which are fully determined in terms of a reduce density matrix $\widehat\rho_\sys(\tau)$ for the system modes $\mf{k}\in\sys$,
\begin{equation}
    \widehat\rho_\sys(\tau)
    \equiv\Tr_{\mf{k}\notin \sys}[\ket{\Psi(\tau)}\bra{\Psi(\tau)}]
    \,.
\end{equation}
{Here, $\tau$ denotes the conformal time, which is related to the scale factor $a(\tau)$ and the Hubble parameter $H$ as $\tau\simeq-(a(\tau) H)^{-1}$.}
The reduced density matrix is useful for quantifying the remaining quantum coherence in the system. For instance, the purity $\mathcal{P}_\sys\equiv \Tr_{\mf{k} \in\sys}[\widehat\rho_\sys^2]$ serves as a basis-independent measure; $\mathcal{P}_\sys=1$ for the pure state, while $(0\leq) \mathcal{P}_\sys\ll1$ for the system far from being pure. It would be hard to prove the quantumness of $\zeta$ by measuring the sector $\sys$ alone if $\mathcal{P}_\sys\ll1$. 

As an illustrative example, we choose a pair of superhorizon modes $(\mf{q},-\mf{q})$ that satisfy $|\mf{q}|\equiv q\ll a(\tau_\text{e})H\}$ as the system following the previous literature, except in Section~\ref{sec:IRdiv}. Here, $\tau_\text{e}$ denotes the end of inflation time.
Other modes are referred to as environment modes. 
Note that it is of course possible to consider other choices: for instance, one may take all modes satisfying $k=q$ as system variables. We expect that this will not change the main part of our results.

It is useful to start with the theory in a finite box of the comoving volume $V$ and take the infinite box size limit later. The Fourier modes are discretized as $k_i(n_i)=n_i\pi/L_i$ for $i=x,y,z$ with arbitrary set of integers $\{n_i\}$, where $\{L_i\}$ denotes the size of the box in the $i$-direction. We then choose the pair of superhorizon modes $(\mf{k}(\mf{n}_\mf{q}),-\mf{k}(\mf{n}_\mf{q}))$ as the system variables, where $\lim_{\{L_i\}\to\infty}\mf{k}(\mf{n}_\mf{q})=\mf{q}$. Below, we write the discrete sum $\sum_{\mf n}$ as the three-dimensional integration of continuous momenta assuming that we take the infinite volume limit, while keeping the $V$-dependence in equations.

The loss of coherence is caused by non-linear couplings between the system modes and the environment modes $\mf{k}$ with $k\neq q$. Such couplings lead to non-zero values of higher-order wavefunction coefficients $\psi_n$ with $n\geq 3$, which parameterize the wavefunction in the field basis as
\begin{equation}
    \Psi[\zeta;\tau] = 
    \mc{N}
    \exp\Biggl[
    -\sum_{n=2}^\infty \frac{1}{n!}
    \left(
    \prod_{j=0}^n\int_{\mf{k}_j}\,\zeta_{\mf{k}_j}(\tau)
    \right)
    \psi_n(\{\mf{k}_i\};\tau)
    (2\pi)^3\delta^{(3)}\Big(\sum_i\mf{k}_i\Big)
    \Biggr]
    \,,
    \label{eq:wfu}
\end{equation}
where $\int_\mf{k}\equiv(2\pi)^{-3}\int \d^3\mf{k}$.
Since the theory is weakly coupled, we trace out the environment perturbatively. Focusing on the Gaussian pieces of $\rho_\sys$, we obtain\footnote{To avoid the double counting, we treated $\Re\zeta_\mf{q}$ and $\Im\zeta_\mf{q}$ as independent variables, whereas $\zeta_\mf{-q}$ is no longer independent and completely fixed by the condition $\zeta_\mf{-q}=\zeta_\mf{q}^*$. As a result, we have the following decomposition; $\sum_{\mf{n}} |\zeta_\mf{n}(\tau)|^2=2|\zeta_\mf{n_q}(\tau)|^2 + \sum_{\mf{n}\neq \pm\mf{n_q}}|\zeta_\mf{n}(\tau)|^2$. This is the reason why we do not have a factor $1/2$ in front of $A_q$ and $A_q^*$ in the exponent.}
\begin{align}
    \eval{\rho_\sys[\zeta_\mf{q},\widetilde\zeta_\mf{q};\tau]}_\text{Gaussian}
    =
    \frac{\Re A_q(\tau){-}C_q(\tau)}{V\pi}\,
    e^{-\frac{A_q(\tau)}{V}|\zeta_\mf{q}|^2-\frac{A^*_q(\tau)}{V}|\widetilde{\zeta}_\mf{q}|^2}e^{\frac{C_q(\tau)}{2V}(\zeta_\mf{q}\widetilde{\zeta}^*_{\mf{q}}+\zeta^*_{\mf{q}}\widetilde{\zeta}_\mf{q})}
    \,,\label{rho_G_eg}
\end{align}
where we used $\zeta_{-\mf{q}}=\zeta^*_\mf{q}$. We assume $C<\Re A$ to ensure that the state is normalizable. We have $A_q(\tau)=\psi_{2}(\mf{q};\tau)$ and $C_q(\tau)=0$ in a free theory.

The last term of \eqref{rho_G_eg} denotes the non-separable piece; the state is mixed when this term is present. This piece is generated at one-loop level as (see e.g.,~\cite{Sou:2022nsd})
\begin{align}
    \eval{C_q(\tau)}_\text{one-loop}
    =
    \frac{1}{2}\int_{\bf k}\int_{\bf k'} P_k(\tau)P_{k'}(\tau)|\psi_3(\mf{q}, \mf{k}, \mf{k}')|^2(2\pi)^3\delta^{(3)}(\mf{q}+\mf{k}+\mf{k}')
    \,, \label{formula}
\end{align}
{where $P_k(\tau)$ denotes the tree-level power spectrum of $\zeta$; $P_k(\tau)(2\pi)^3\delta^{(3)}(\mf{k} +\mf{k}')=\langle\zeta_\mf{k}(\tau)\zeta_\mf{k'}(\tau)\rangle_\text{tree}$.} This shows the positivity $C>0$.
Indeed, the purity is unity only when $C=0$, that is, the full system is separable;
\begin{align}
    \mathcal{P}_\sys
    =
    \int\mathrm{d}^2\zeta_{\mf q}\mathrm{d}^2\widetilde{\zeta}_{\mf q}\,
    | \rho_\sys[\zeta_{\mf q}, \widetilde{\zeta}_{\mf q};\tau]|
    ^2
    \simeq 
    \frac{\Re A{-}C}{\Re A{+}C}
    \,,\label{purity_rhoG}
\end{align}
where  $\int\mathrm{d}^2X_{\mf q}\equiv \frac{1}{2}\int^\infty_{-\infty}\mathrm{d}(\Re X_\mf{q})\int^\infty_{-\infty}\mathrm{d}(\Im X_\mf{q})$ for 
$X=\zeta,\widetilde{\zeta}$, and we used the Gaussian approximation for $\rho_\sys$ in the second line.

The formula \eqref{formula} enables us to compute the non-separable piece of $\rho_\sys$ at the one-loop level by evaluating the tree-level cubic coupling $\psi_3$, analogous to the on-shell factorization of scattering amplitudes in flat space. Note that Eq.~\eqref{formula} will apply to more generic setup, including weakly coupled theories in flat space.

\subsection{Consistency relation and divergent decoherence rate}\label{sec:div1}

\subsubsection{Consistency relation for wavefunction coefficients}
The Maldacena's consistency relation \cite{Maldacena:2002vr} can be understood as one of the consequences of the Ward identity following from invariance of the wavefunction under the spatial diffeomorphism $x^i \mapsto x^i+v^i$ \cite{Pimentel:2013gza}. Imposing the invariance $\Psi[h_{ij}+D_{(i}v_{j)}] =\Psi[h_{ij}]$, where $D_i$ denotes the three-dimensional covariant derivative, we obtain
\begin{align}
    \lim_{k_1\to0}\psi_3(\mf{k}_1,\mf{k}_2,\mf{k}_3)=\qty(3-k_2\pdif{k_2})\psi_2
    \label{consistency}
\end{align}
as the relation between $\psi_2$ and $\psi_3$ in the squeezed limit.\footnote{Strictly speaking, \eqref{consistency} is derived under the additional assumption that the behavior of $\psi_3$ in the squeezed limit $k_1\to0$ is milder than $\log k_1$, which may be related to a condition of locality \cite{Berezhiani:2013ewa,Tanaka:2017nff}. This assumption will be valid at least at tree level in the present setup, which is sufficient for computing purity at one-loop level.}
In this section, we focus on the decoherence rate induced by the deep-UV or deep-IR modes for which we can use the relation \eqref{consistency}. We will revisit the other contributions in Section~\ref{sec:finite_deco}. 
It is worth noting that the leading contributions to 
$|\psi_3|$ in the equilateral configurations come from the cubic boundary term $\del_t(9\Mpl^2 H a^3\zeta^3)$, which is in fact correctly reproduced from \eqref{consistency} as you can check in the next paragraph. This is because this interaction is independent of $k$.

The wavefunction coefficients can be calculated using the semiclassical approximation
\begin{align}
    \Psi[\zeta]\approx e^{iS_\text{cl}[\zeta]}
    \,,
\end{align}
{at the tree level.}
The expression of $\psi_2(k)$ is derived in Appendix \ref{sec:psi2}. Here we cite the result:
\begin{align}
    \psi_2=\frac{2i\mpl^2k^3}{H^2} \qty(\epsilon\qty(\frac{aH}{k})^2\frac{H_{\nu-1}^{(2)}(-k\tau)}{H_\nu^{(2)}(-k\tau)}+9\qty(\frac{aH}{k})^3-\frac{aH}{k}),
    \label{eq:psi2}
\end{align}
where $\epsilon=-\dot{H}/H^2$ is the slow-roll parameter. This expression is valid up to second order in slow-roll.
{Note that the first term comes from bulk terms, so we denote it as $\psi_{2,\text{bulk}}$ for notational simplicity. Similarly, the second and third terms are from boundary terms, denoted as $\psi_{2,\text{bd}}$.}

In superhorizon and subhorizon limit,
\begin{align}
    \psi_2&\underset{k\to 0}{\longrightarrow}\frac{2\mpl^2k^3}{H^2}\qty(\epsilon\qty(\qty(\frac{k}{2aH})^{1-n_s}+\gamma(1-n_s))+9i\qty(\frac{aH}{k})^3-i\frac{aH}{k}(1-\epsilon +3\epsilon^2+\epsilon\eta)),\\
    \psi_2&\underset{k\to \infty}{\longrightarrow}\frac{2\mpl^2k^3}{H^2}\qty(\epsilon\qty(\frac{aH}{k})^2+i\qty(\frac{aH}{k})^3\qty(9+\epsilon-2\epsilon^2+\frac{\epsilon\eta}{2})-i\frac{aH}{k}),
\end{align}
where $\gamma=(\pi i+2(2\ln{2}-2+\gamma_\text{E}))/2$ with $\gamma_\text{E}$ being Euler--Mascheroni constant, $\eta=\dot{\epsilon}/H\epsilon$, and $n_s=1-2\epsilon-\eta$.
By applying the consistency relation to Eq.~\eqref{eq:psi2}, the expression of $\psi_3$ reads 
\begin{align}
    \psi_3\underset{k_1\to 0}{\longrightarrow}\frac{2i\mpl^2k_2^3}{H^2}\Bigg({\epsilon(n_s-1)}\qty(\frac{aH}{k_2})^2\frac{H_{\nu-1}^{(2)}(-k_2\tau)}{H_\nu^{(2)}(-k_2\tau)}&+\epsilon(1-\epsilon)\frac{aH}{k_2}\Bigg(\frac{H_{\nu-1}^{(2)}(-k_2\tau)}{H_\nu^{(2)}(-k_2\tau)}\Bigg)^2\notag\\
    &+27\qty(\frac{aH}{k_2})^3-{(1-\epsilon)}\frac{aH}{k_2}\Bigg).
    \label{3zetasoft}
\end{align}
For the situation where both of a long mode and short modes are in the superhorizon,
\begin{align}
    \psi_3\underset{|k_1\tau|\ll |k_2\tau|\ll 1}{\longrightarrow}\frac{2\mpl^2k_2^3}{H^2}\qty({\epsilon(n_s-1)}+27i\qty(\frac{aH}{k_2})^3-i\frac{aH}{k_2}{(1-\epsilon+3\epsilon^2+\epsilon\eta)})
    \label{eq:psi3_superh}
\end{align}
is satisfied. The first term is from $\Re\psi_{2,\text{bulk}}$, and reproduces the Maldacena's consistency relation. Also, the terms of order ${\epsilon^0}$ and ${\epsilon}$ in the second and third terms would correspond to the cubic boundary terms $\partial_\tau(9a^3\mpl^2 H \zeta^3)$ and $\partial_\tau((1-\epsilon)a\zeta(\partial_i\zeta)^2/H)$. Note that the terms of order ${\epsilon^0}$ are from $\psi_{2,\text{bd}}$, but the terms of order ${\epsilon}$ and ${\epsilon^2}$ are from $\psi_{2,\text{bulk}}$.
The case of short modes being subhorizon leads to
\begin{align}
    \psi_3\underset{|k_1\tau|\ll 1\ll |k_2\tau|}{\longrightarrow}\frac{2\mpl^2k_2^3}{H^2}\qty(2\epsilon\qty(\frac{aH}{k_2})^2 +i\qty(\frac{aH}{k_2})^3\qty(27+3\epsilon-6\epsilon^2+\frac{3\epsilon\eta}{2}) -i\frac{aH}{k_2}
    ),
    \label{eq:psi3_subh}
\end{align}
where we kept terms up to $\order{(k_2\tau)^{-3}}$.

\subsubsection{IR and UV divergences}

Using Eqs.~\eqref{formula}, \eqref{eq:psi2}--\eqref{eq:psi3_subh}, the IR and UV contributions are calculated as follows.
Firstly for IR contributions,
\begin{align}
    C_{q,\text{IR}}\approx \frac{H^4}{32\pi^2\epsilon^2\mpl^4}\frac{|\psi_{3,\mr{IR}}(q)|^2}{q^3}\int_{k\sim k_\text{min}} \frac{k^2\d{k}}{k^3}\propto \log{k_\text{min}},
\end{align}
where $\psi_{3,\mr{IR}}$ is given in \eqref{eq:psi3_superh}, and the IR divergence is from the long mode propagator.
For UV contributions,
\begin{align}
    C_{q,\text{UV}}\approx \frac{1}{16\pi^2\epsilon^2}\int_{k\sim a\Lambda} k^2\d{k}&\Bigg(4\epsilon^2+\qty(\frac{k}{aH})^2-2\qty(27+3\epsilon-6\epsilon^2+\frac{3\epsilon\eta}{2})\notag\\
    &\qquad\qquad +\qty(\frac{aH}{k})^2\qty(27+3\epsilon-6\epsilon^2+\frac{3\epsilon\eta}{2})^2\Bigg),
\end{align}
which is badly UV divergent, namely $\Lambda^3$ for bulk interactions and $\Lambda^5$ for boundary ones.

\subsection{Growing terms, perturbativity, and late time resummation}
We see that $C$ is positive and grows at a late time.
If there were no such growing terms in $A$, this would imply the large loop correction to the power spectrum given by $(\Re A - C)^{-1}$; e.g., the one-loop correction is 
\begin{align}
    \eval{P_\zeta(q;\tau)}_\text{one-loop}
    =
    \frac{-\Re A_{q,\text{one-loop}}(\tau) +C_{q,\text{one-loop}}(\tau)}{(\Re A_{q,\text{tree}}(\tau))^2}
    \,,
\end{align}
which grows at a late time as fast as inverse powers of $\tau$ unless the cancellation occurs between the growing terms in $\Re A$ and $-C$. 

\begin{figure}[t]
\centering
\begin{fmffile}{bulk_correction}
\begin{align*}
\begin{fmfgraph*}(100,40)
\fmfleft{i0,i1}
\fmfright{o0,o1}
\fmf{plain,tension=1.3}{i1,v1}
\fmf{plain,tension=1.3}{v2,o1}
\fmf{plain,left=1,tension=1.5}{v1,v2,v1}
\fmf{phantom,tension=2}{i0,v1}
\fmf{phantom,tension=2}{v2,o0}
\begin{tikzpicture}[overlay]
     \draw[densely dotted, line width=0.9pt] (0,1.4) -- (3.5,1.4);
     \node[left] at (0,1.4) {$\tau$};
\end{tikzpicture}
\end{fmfgraph*}
\hspace{2.5cm}
\begin{fmfgraph*}(100,40)
\fmfleft{i0,i1}
\fmfright{o0,o1}
\fmf{plain,tension=0.5}{i1,v1}
\fmf{plain,tension=0.5}{v1,o1}
\fmf{plain,tension=1}{v1,v2}
\fmf{plain,left=1,tension=0.2}{v2,v3,v2}
\fmf{phantom,tension=3}{i0,v3}
\fmf{phantom,tension=3}{v3,o0}
\begin{tikzpicture}[overlay]
     \draw[densely dotted, line width=0.9pt] (0,1.4) -- (3.5,1.4);
     \node[left] at (0,1.4) {$\tau$};
\end{tikzpicture}
\end{fmfgraph*}
\end{align*}
\end{fmffile}
\caption{Corrections from the quantum fluctuations around the semi-classical time evolution in the bulk. A bulk-to-boundary propagator is assigned for the lines connected with the future boundary. A bulk-to-bulk propagator is assigned for the internal lines connecting two bulk points.}
\label{fig:bulk}
\end{figure}
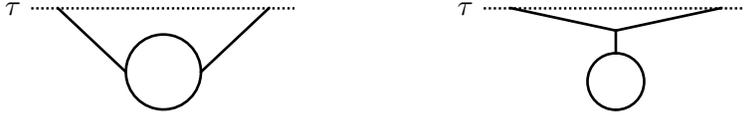

{To see the cancellation,} we first focus on the contributions from superhorizon environment {around some comoving scale $k_*$ with $q\ll k_*$.} In this case, it is convenient to separate the one-loop corrections to $A$ from cubic vertices into two pieces: the correction due to the quantum fluctuations of the environment at the final-time slice, and the contribution from the quantum corrections to the bulk time evolution. The latter is taken into account as the correction to the wavefunction coefficient $\psi_2$. We refer to these corrections as $A_\text{one-loop,bd}$ and $A_\text{one-loop,bulk}$, respectively.

The former contribution is evaluated as 
\begin{align}
    A_\text{one-loop,bd}
    =
    -\frac{1}{2}
    \int_{\bf k}\int_{\bf k'} 
    P_k(\tau)P_{k'}(\tau)\psi_3^2(\mf{q}, \mf{k}, \mf{k}';\tau)(2\pi)^3\delta^{(3)}(\mf{q}+\mf{k}+\mf{k}')
    \,,\label{Abd}
\end{align}
which grows as fast as inverse powers of $\tau$ according to \eqref{eq:psi3_superh}. 
Another term $A_\text{one-loop,bulk}$ 
receives contributions from two diagrams $(a)$ and $(b)$ shown in Fig.~\ref{fig:bulk}. 
The bulk-to-bulk propagator $G(\tau',\tau'';k)$ is assigned to the line connecting two bulk points, and the bulk-to-boundary propagator $K_k(\tau')$ is assigned to the line connecting a single bulk point to the future boundary. The former is related to the Feynman propagator $G_{++}(\tau',\tau'';k)=\langle T[\zeta_\mf{k}(\tau)\zeta_\mf{k'}(\tau')]\rangle'$ as  $G(\tau',\tau'';k)=G_{++}(\tau',\tau'';k)-P_k(\tau)K_k(\tau')K_k(\tau'' )$, while the latter is $K_k(\tau')={u_k^\zeta}^*(\tau')/{u_k^\zeta}^*(\tau)$. Here, $T[\cdots]$ denotes the time-ordered products, and $u_k^\zeta(\tau)$ denotes the mode function of $\zeta_\mf{k}$.
Since $G$ decays at superhorizon scales as $G(\tau',\tau'';k)= \mathcal{O}(|\tau'|^3, |\tau''|^3)$, we find by straightforward computation that $A_\text{one-loop,bulk}$ can exhibit at most the logarithmic growth and is much smaller than the leading term in $A_\text{one-loop,bd}$. 
Physically, this sub-dominance of $A_\text{one-loop,bulk}$ reflects the fact that the time evolution of each superhorizon mode is well approximated by the classical physics after the horizon crossing. We expect this to remain valid even after the end of inflation, since the bulk-to-bulk propagator remains suppressed relative to the bulk-to-boundary propagator at superhorizon scales.

Hence, as long as superhorizon environment is concerned, the dominant piece in $A_\text{one-loop}$ is determined by the boundary contribution \eqref{Abd}. We now compute $\Re A-C$ at one-loop level only keeping the terms which grows as fast as inverse powers of $\tau$ as
\begin{align}
    \eval{[\Re A-C]_\text{one-loop}}_{k\sim k_*}
     &\simeq
     -\frac{1}{4}\int_{\mf{k},{k\sim k_*}}P_kP_{k'}(\psi_3^2+{\psi^*_3}^2+2|\psi_3|^2)
    =
    -\int_{\mf{k},{k\sim k_*}}P_kP_{k'}(\Re\psi_3)^2
    \,,
\end{align}
which does not grow at a late time according to \eqref{eq:psi3_superh}. Here, we abbreviated the arguments of $\psi_3$. 

\begin{figure}[t]
\centering
\begin{fmffile}{configuration}
\begin{fmfgraph*}(150,35)
\fmfleft{i0,i1}
\fmfright{o0,o1}
\fmf{plain,tension=1.5}{i1,v1}
\fmf{plain,tension=1.5}{v2,o1}
\fmf{plain,left=1,label=$G_{\text{adv}}$}{v1,v2}
\fmf{plain,left=1,label=$G_{\text{ret}}$}{v2,v1}
\fmflabel{$\tau'$}{v1}
\fmflabel{$\tau''$}{v2}
\end{fmfgraph*}
\end{fmffile}
\caption{A one-loop diagram where the internal lines are the retarded and advanced Green's function ($G_\text{ret}(\tau',\tau'')$ and $G_\text{adv}(\tau',\tau'')$), respectively. This diagram is exactly zero since the product of these two Green's functions vanishes, which explains the cancellation of the dominant pieces in $\Re A$ and $C$; see Appendix~\ref{sec:nogrow} for more details.
}
\label{fig:cc}
\end{figure}
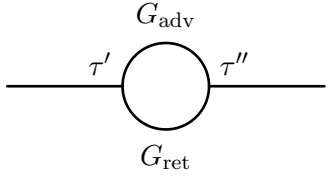

{To see the cancellation including the contributions from all modes shorter than $q$,} the bulk term $A_\text{one-loop,bulk}$ is no longer negligible. A more convenient approach to demonstrate the cancellation of growing terms is to use the Schwinger--Keldysh formalism. Further details are provided in Appendix~\ref{sec:nogrow}, which shows that the cancellation occurs because the dominant components of $\Re A$ and $-C$ combine to form the diagram in Fig.~\ref{fig:cc}, which vanishes since it is given by the integral of the product of the retarded and advanced Green’s functions.

Hence, the presence of growing term in $\Re A$ and $C$ is consistent with the perturbativity of correlation functions, and also leads to the condition $\Re A_\text{one-loop}\simeq C_\text{one-loop}$ at the one-loop level.

This condition is practically useful for evaluating the late-time purity. At the early time when the one-loop correction to purity is small, we can perform the perturbative expansion in \eqref{purity_rhoG}, yielding
\begin{align}
    \mathcal{P}_\sys
    \simeq 
    1 - \frac{2C_\text{one-loop}}{\Re A_\text{tree}}
    =
    1 - 4P_qC_\text{one-loop}
    \,.\label{purity_earl}
\end{align}
Here, $P_q$ denotes the tree power spectrum of $\zeta_\mf{q}$. 
This expansion is not valid at a late time due to the growth of $C$, and it is necessary to consider \eqref{purity_rhoG} which is exact under the Gaussian approximation. Using $\Re A_\text{one-loop}\simeq C_\text{one-loop}$, we obtain
\begin{align}
    \mathcal{P}_\sys
    \simeq
    \frac{1}{1+\Delta_q(\tau)}
    \,,\qquad
    \Delta_q(\tau)
    =
    4P_qC_\text{one-loop}
    \,.\label{purity_late}
\end{align}
Therefore, we can easily obtain the resummed form by evaluating the perturbative correction to the purity via \eqref{purity_earl}.

\section{Effective quantum state for an observer}\label{sec:eff_state}
In the previous section, we found IR and UV divergences in the one-loop reduced density matrix $\rho_\sys$. 
As we will see in Sections~\ref{sec:IRdiv} and \ref{sec:UVdiv}, however, these divergences are absent in the correlation functions of properly defined curvature perturbations which resemble what a local observer can actually measure using a detector with finite time resolution.
More details will be discussed in these sections, but the punchline is that the equal-time correlators of the original variables $\zeta(\tau)$ and $\pi(\tau)$, which can be evaluated using $\rho_\sys$, are not appropriate observables at loop level.

Let us denote the properly defined curvature perturbations and its conjugate momenta as $\overline{\zeta}^g_\mf{\pm q}$ and $\overline{\pi}^g_\mf{\pm q}$, respectively. What does the quantum state look like to an observer who is restricted to measuring the correlation functions of these operators? We propose to define quantum states based on correlation functions that are accessible to a given observer.

\subsection{Preliminary: Wigner function}
For convenience, we introduce the formulation base on the Wigner function $W[\zeta,\pi]$ which is related to $\rho_\sys$ via the Wigner--Weyl transform;
\begin{align}
    W\left[\frac{\zeta_\mf{q}+\widetilde{\zeta}_\mf{q}}{2}, \pi_\mf{q};\tau\right]
    &\equiv
    \int\mathrm{d}^2(\zeta_\mf{q}-\widetilde{\zeta}_\mf{q})\,
    \rho_\sys[\zeta_\mf{q}, \widetilde{\zeta}_\mf{q};\tau] 
    e^{
    -\frac{i}{V}
    \left[
        \pi_\mf{q}(\zeta^*_\mf{q}-\widetilde{\zeta}^*_\mf{q})
        +
        \pi^*_\mf{q}
        (\zeta_\mf{q}-\widetilde{\zeta}_\mf{q})
    \right]
    }
    \,.\label{wig_def}
\end{align}
The normalization condition is
\begin{align}
    \int\mathrm{d}^2\zeta_\mf {q}\int\frac{\mathrm{d}^2\pi_\mf{q}}{(2\pi)^2}\,
    W[\zeta_\mf{q},\pi_\mf{q};\tau]
    = V^2\,,
\end{align}
where the volume factor appears since the left-hand side (LHS) represents the probability density in the momentum space. 
Since $\pi_\mf{q}$ is conjugate to the off-diagonal element of $\rho_\sys$, the Wigner function serves as a (pseudo) weight function which gives the expectation values of $\zeta$ and $\pi$ written in the Weyl order. 

The Wigner function under the Gaussian approximation is therefore
\begin{align}
    W[\zeta_\mf{q},\pi_\mf{q};\tau] 
    \simeq 
    \frac{1}{\det\cov} 
    \exp\left[-\frac{1}{2V}\Re \Phi^\alpha(\cov^{-1})_{\alpha\beta}\Re \Phi^\beta 
    \right]
    \exp\left[-\frac{1}{2V}\Im \Phi^\alpha(\cov^{-1})_{\alpha\beta}\Im \Phi^\beta 
    \right]
    \,,\label{wigner_G}
\end{align}
where the indices $\alpha$ and $\beta$ run over the set $\{\zeta,\pi\}$, and $\cov^{-1}$ represents the inverse of the $2 \times 2$ covariance matrix $\cov$, defined as $\cov_{\alpha\beta} \equiv \langle \{\Phi^\alpha_\mf{q}(\tau), \Phi^\beta_\mf{q'}(\tau)\} \rangle'$ with $(\Phi^\zeta_\mf{q}, \Phi^\pi_\mf{q})=(\zeta_\mf{q}, \pi_\mf{q})$.
In addition, $\{X,Y\}\equiv \frac{1}{2}(XY+YX)$ and $\langle\cdots\rangle'$ denote the correlator in the momentum space divided by the delta function $(2\pi)^3\delta(\mf{q+q'})$; in this formula, the factor $1/V$ in the exponent accounts for the divergence of this delta function at $\mf{q}=-\mf{q}'$, and the momentum conservation is ensured by the absence of cross-correlation between $\Re\Phi^\alpha$ and $\Im\Phi^\alpha$.

For the Gaussian state \eqref{rho_G_eg}, the covariance matrix is 
\begin{align}
    \cov
    =
    \frac{1}{2(\Re A-C)}
    \begin{pmatrix}
        1 & -\Im A \\
        -\Im A & |A|^2-C^2
    \end{pmatrix}
    \,.
    \label{cov_G}
\end{align}
The purity for the Gaussian Wigner function is
\begin{align}
    \mathcal{P}_\sys
    =
     V^{-2}
     \int\mathrm{d}^2\zeta_\mf {q}\int\frac{\mathrm{d}^2\pi_\mf{q}}{(2\pi)^2}\,
     \left(W[\zeta_\mf{q},\pi_\mf{q};\tau]\right)^2
     \simeq
     \frac{1}{4\det\cov}
    \,.\label{purity_wigG}
\end{align}
Eqs.~\eqref{cov_G} and \eqref{purity_wigG} correctly reproduce the purity \eqref{purity_rhoG} evaluated using the reduced density matrix.
At the tree-level, the purity becomes unity since the condition $\det\cov=1/4$ is ensured by the Wronskian 
\begin{align}
    u_q^\pi(\tau){u_q^{\zeta*}}(\tau)-u_q^\zeta(\tau){u_q^{\pi*}}(\tau)
    =-i\,,\label{wronskian}
\end{align}
where $u_q^{\zeta}$ and $u_q^{\pi}$ are the mode functions of $\zeta$ and $\pi$ in the interaction picture, respectively. Eq.~\eqref{wronskian} also ensures the canonical commutation relation at the tree level. 
Note that the non-negativity of $\mathcal{P}_\sys$ is automatically ensured by the reality of $W$
, which is equivalent to the condition $\rho_\sys[\zeta_\mf{q},\widetilde{\zeta}_\mf{q};\tau]=\rho_\sys^*[\widetilde{\zeta}_\mf{q},\zeta_\mf{q};\tau]$.

\subsection{Effective quantum state}
To define effective quantum state based on correlation functions, it is useful to construct quantum state from generating functional.\footnote{This treatment is analogous to the Gram--Charlier and Edgeworth series for random variables.}
The connection between equal-time correlation functions and the quantum state at the time $t$ can be clarified using the generating function of equal-time Weyl-ordered correlators, 
\begin{align}
    Z_\text{w}(J^\zeta_\mf{q},J^\pi_\mf{q};\tau)
    \equiv
    &\int\mathscr{D}\zeta_\pm\int\mathscr{D}\pi_\pm\,
    \delta(\zeta_\Delta(\tau))
    e^{
        i\left[S_\text{H}[\zeta_+,\pi_+]-S_\text{H}[\zeta_-,\pi_-]\right]
    }
    e^{\frac{i}{V}\left[\left(J^\zeta_\mf{q}\zeta^*_{\mf{q},c}(\tau)+J^\pi_\mf{q}\pi^*_{\mf{q},c}(\tau)\right) + \text{(c.c.)}\right]}
    \,,\label{Z_def}
\end{align}
with 
\begin{align}
    \zeta_\Delta(\tau)\equiv \zeta_+(\tau)-\zeta_-(\tau)
    \,,\quad
    \zeta_{\mf{q},c}(\tau)\equiv \frac{\zeta_\mf{q,+}(\tau)+\zeta_\mf{q,-}(\tau)}{2}\,,\quad
    \pi_{\mf{q},c}(\tau)\equiv \frac{\pi_\mf{q,+}(\tau)+\pi_\mf{q,-}(\tau)}{2}
    \,.
\end{align}
Here, $\mathscr{D}\zeta_\pm$ and $\mathscr{D}\pi_\pm$ are the usual path integration over the $\pm$ variables $(\zeta_+,\zeta_-)$ and $(\pi_+,\pi_-)$, respectively, and $S_\text{H}$ denotes the action written in terms of the Hamiltonian, i.e., $S_\text{H}=\int(\pi\dot{\zeta}-H)$. 
We also employed the $i\epsilon$-prescription and omit the dependence on the initial wavefunctions. The  reduced density matrix and the Wigner function are then given by 
\begin{align}
    &\rho_\sys[\zeta_\mf{q},\widetilde{\zeta}_\mf{q};\tau]
    =
    \int\frac{\mathrm{d}^2J^\zeta_\mf{q}}{(2\pi V)^2}\,
    Z_\text{w}[J^\zeta_\mf{q},\zeta_\mf{q}-\widetilde{\zeta}_\mf{q};\tau]
    \,e^{-\frac{i}{V}\left[J^\zeta_\mf{q}\left(\frac{\zeta^*_\mf{q}+\widetilde{\zeta}^*_\mf{q}}{2}\right)+\text{(c.c)}\right]}
    \,,\label{rho_Z}\\
    &W[\zeta_\mf{q},\pi_\mf{q};\tau]
    =
    \int\frac{\mathrm{d}^2J^\zeta_\mf{q}}{(2\pi V)^2}
    \int\mathrm{d}^2J^\pi_\mf{q}\,
    Z_\text{w}[J^\zeta_\mf{q},J^\pi_\mf{q};\tau]
    \,e^{-\frac{i}{V}\left[J^\zeta_\mf{q}\zeta^*_\mf{q}+\text{(c.c)}\right]}
    e^{-\frac{i}{V}\left[J^\pi_\mf{q}\pi^*_\mf{q}+\text{(c.c)}\right]}
    \,.\label{wig_Z}
\end{align} 
Eqs.~\eqref{rho_Z} and \eqref{wig_Z} with \eqref{Z_def} are consistent with \eqref{wig_def}. 

It is straightforward to construct the effective generating function $\overline{Z}_\text{w}^g$ as
\begin{align}
    \overline{Z}_\text{w}^g(J^\zeta_\mf{q},J^\pi_\mf{q};\tau)
    =
    \int\mathscr{D}\zeta_\pm\int\mathscr{D}\pi_\pm\,
    \delta(\zeta_\Delta(\tau))
    e^{
        i\left[S_\text{H}[\zeta_+,\pi_+]-S_\text{H}[\zeta_-,\pi_-]\right]
    }
    e^{\frac{i}{V}\left[\left(J^\zeta_\mf{q}\overline{\zeta}^{g\,*}_{\mf{q},c}(\tau)+J^\pi_\mf{q}\overline{\pi}^{g\,*}_{\mf{q},c}(\tau)\right) + \text{(c.c.)}\right]}
    \,,
\end{align}
from which one can evaluate only the correlation functions of $\overline{\zeta}^g_\mf{\pm q}$ and $\overline\pi^g_{\pm q}$. In terms of $\overline{Z}_\text{w}^g$, we define the effective $\rho_\sys$ and $W$ as 
\begin{align}
    &\overline{\rho}_\sys^g[\zeta_\mf{q},\widetilde{\zeta}_\mf{q};\tau]
    \equiv
    \int\frac{\mathrm{d}^2J^\zeta_\mf{q}}{2\pi}\,
    \overline{Z}_\text{w}^g[J^\zeta_\mf{q},\zeta_\mf{q}-\widetilde{\zeta}_\mf{q};\tau]
    \,e^{-\frac{i}{V}\left[J^\zeta_\mf{q}\left(\frac{\zeta^*_\mf{q}+\widetilde{\zeta}^*_\mf{q}}{2}\right)+\text{(c.c)}\right]}
    \,,\label{Srho_Z}\\
    &\overline W^g[\zeta_\mf{q},\pi_\mf{q};\tau]
    \equiv
    \int\frac{\mathrm{d}^2J^\zeta_\mf{q}}{2\pi}\int\frac{\mathrm{d}^2J^\pi_\mf{q}}{2\pi}\,
    \overline{Z}_\text{w}^g[J^\zeta_\mf{q},J^\pi_\mf{q};\tau]
    \,e^{-\frac{i}{V}\left[J^\zeta_\mf{q}\zeta^*_\mf{q}+\text{(c.c)}\right]}
    e^{-\frac{i}{V}\left[J^\pi_\mf{q}\pi^*_\mf{q}+\text{(c.c)}\right]}
    \,,\label{Swig_Z}
\end{align}
which are obtained from \eqref{rho_Z} and \eqref{wig_Z} by just replacing $Z_\text{w}$ by $\overline{Z}_\text{w}^g$. We can check that $\overline{\rho}_\sys^g$ and $\overline{W}^g$ are related via the Wigner--Weyl transformation as in the usual case \eqref{wig_def}.

Under the Gaussian approximation, $\overline{W}^g$ is given by the effective covariance matrix $\overline{\cov}^g$, each entry of which is a two-point function of $\overline{\zeta}^g$ and $\overline{\pi}^g$.
\begin{align}
    \overline W^g[\zeta_\mf{q},\pi_\mf{q};\tau]
    \simeq 
    \eval{\eqref{wigner_G}}_{\cov\to\overline\cov}
    \,,\,\qquad
    \overline\cov^g_{\alpha\beta}\equiv \langle\{\overline\Phi^{g\,\alpha}_{\bf q}(\tau),\overline\Phi^{g\,\beta}_{\bf q'}(\tau)\}\rangle'
    \,.
    \label{Ewigner_G}
\end{align}
The effective reduced density matrix $\overline{\rho}^g_\sys$ is obtained from $\overline{W}^g$ via the Wigner--Weyl transformation. Under the Gaussian approximation, $\overline{\rho}^g_\sys$ takes exactly the same form as the original expression \eqref{rho_G_eg}, but with the coefficients $A$ and $C$ replaced by the corresponding coefficients $\overline{A}^g$ and $\overline{C}^g$, which reproduce $\overline{\cov}^g$ correctly according to the formula $\eqref{cov_G}|_{(A,C) \mapsto (\overline{A}^g,\overline{C}^g)}$.

For illustration, let us consider measuring the expectation value of some {system} operator $\widehat B$, whose representation in the $\zeta$-basis is $B[\zeta,\widetilde\zeta]$. Then, its expectation value is theoretically given by 
\begin{align}
    \langle\widehat B\rangle_\text{theory}
    =
    \int\mathrm{d}\zeta\int\mathrm{d}\widetilde\zeta\,
    B[\zeta,\widetilde\zeta]\rho_\sys[\widetilde\zeta,\zeta]
    \,.\label{theory}
\end{align}
A practical way of measuring the expectation value of $\widehat{B}$ will be to reconstruct $\rho_\sys[\widetilde\zeta,\zeta]$ from the correlators of $\zeta$ and $\pi$ and substitute the result into \eqref{theory}.
However, a local observer who measures only correlators of $\overline\zeta^g$ and $\overline\pi^g$ will reconstruct the effective reduced density matrix $\overline\rho_\sys^g[\widetilde\zeta,\zeta]$ instead of the original one. Therefore, an expectation value of $\widehat B$ measured by a local observer will be
\begin{align}
    \langle\widehat B\rangle_\text{local obs}
    =
    \int\mathrm{d}\zeta\int\mathrm{d}\widetilde\zeta\,
    B[\zeta,\widetilde\zeta]\overline\rho_\sys^g[\widetilde\zeta,\zeta]
    \,.\label{obs}
\end{align}

Considering $\widehat{B}$ to be the Gour--Khanna--Mann--Revzen Bell operator~\cite{Gour_2004}, a Bell-type experiment can be designed even for scalar fields (see also Refs.~\cite{Martin:2016tbd,Sou:2024tjv} for cosmological context). 
In particular, Ref.~\cite{Sou:2024tjv} clarified in the context of inflation that the Bell inequality $|\langle\widehat B\rangle_\text{theory}|\leq 2$ is always satisfied when purity is sufficiently tiny.
The above discussion then suggests that, when a local observer performs this experiment, what is actually tested will be an inequality $|\langle\widehat B\rangle|_\text{local obs}\leq 2$. 
Since the only difference between $\overline\rho_\sys^g[\widetilde\zeta,\zeta]$ and the original $\rho_\sys[\widetilde\zeta,\zeta]$ lies in the specific values of coefficients, $\langle\widehat B\rangle_\text{local obs}$ measured by a local observer will be a function of the effective purity $\overline{\mathcal{P}}_\sys^g$ defined by 
\begin{align}
    \overline{P}_\sys^g 
    \equiv 
    \int\mathrm{d}^2\zeta_{\mf q}\int\mathrm{d}^2\widetilde{\zeta}_{\mf q}\,
    | \overline{\rho}_\sys^g[\zeta_{\mf q}, \widetilde{\zeta}_{\mf q};\tau]|
    ^2
    =
    V^{-2}
    \int\mathrm{d}^2\zeta_\mf {q}\int\frac{\mathrm{d}^2\pi_\mf{q}}{(2\pi)^2}\,
    \left(\overline W^g[\zeta_\mf{q},\pi_\mf{q};\tau]\right)^2
    \,,\label{Spurity}
\end{align}
which is analogous to the conventional purity \eqref{purity_rhoG} or \eqref{purity_wigG}. Under the Gaussian approximation, it is simply determined by $\det\overline\cov^g$, just as in \eqref{purity_wigG}:
\begin{align}
    \overline{P}_\sys^g 
    \simeq
    \frac{1}{4\det\overline\cov^g}
    \,.\label{Epurity_G}
\end{align}

A formulation of the Bell inequality for the modified observable $\langle\widehat B\rangle_\text{local obs}$ may be nontrivial in general; however, we expect that it remains valid effectively as long as the effective commutation relation is satisfied in a good approximation. 
In our current setup, commutation relation is satisfied well between $\overline\zeta^g$ and $\overline\pi^g$. Hence, in this study, we assume that the effective purity will serve as a useful probe of quantumness of the system observed by a local observer. It would be interesting to formalize the discussion of Bell-type inequality for this type of effective observables and study the usefulness of effective purity more rigorously, which we leave for future work.

\section{Absorption of IR modes in geodesic coordinate}\label{sec:IRdiv}
\subsection{Formal discussions on IR finiteness}
As mentioned in Section~\ref{sec:basic}, the loop at future time boundary includes logarithmic IR divergence.
One of the treatments for the IR divergence is to rescale the superhorizon constant modes to background metric~\cite{Urakawa:2009my,Urakawa:2010kr,Tanaka:2011aj,Tanaka:2012wi,Tanaka:2013xe,Pajer:2013ana,Tanaka:2013caa,Tanaka:2014ina,Tanaka:2017nff}, leading to local observables described in the geodesic coordinate system. 
A brief review of the treatment is presented in Appendix \ref{sec:loe}, so we just define some quantities here. 
The observer's coordinate, say $x^g$, is characterized by (conformally) free-falling coordinate around origin,
\begin{align}
    g_{\mu\nu}^g=
    \eta_{\mu\nu}a^{2}+\order{R_{\mu\rho \nu\sigma}^gx^{\rho,g}x^{\sigma,g}},
    \label{eq:free_fall}
\end{align}
where the indices $g$ denotes ``geodesic'' since this coordinate $\mf{x}^g$ is actually a geodesic distance. In this paper, we do not discuss 
the detail of the coordinate transformation. Rather practically, we consider $\mf{x}^g\simeq e^{\zeta_\text{L}(\mf{x})}\mf{x}$ where 
$\zeta_{\mathrm{L}}$ {denotes the long-wavelength pieces which we treat as approximately constant over the region of interest,}
\begin{align}
    \zeta_\text{L}(\mf{x})=\int_\mf{k} W_{k_\text{L}}\zeta_\mf{k}e^{i\mf{k}\cdot\mf{x}}
    \label{eq:zeta_L}
\end{align}
with $W_{k_\text{L}}$ being a window function extracting modes longer than a superhorizon scale $k_\text{L}$.\footnote{Note that this effectively amounts to defining geodesic coordinate around the origin using the spatial average $\zeta_\text{long}\equiv\int d^3\mf{x}'W(\mf{x}')\zeta(\mf{x}')$, where $W(\mf{x})$ is a window function for small region around the origin. The use of this averaged value was suggested in \cite{Tanaka:2012wi,Tanaka:2013xe,Tanaka:2014ina} in order to eliminate additional UV divergences reported in \cite{Tsamis:1989yu,Miao:2012xc}.}

Under the coordinate transformation, the curvature perturbations measured in the geodesic coordinate are written as
\begin{align}
    \zeta^g(\mf{x}^g)=
    \zeta(\mf{x}(\mf{x}^g))
    \simeq
    \zeta(\mf{x}^g)+(\mf{x}-\mf{x}^g)\cdot \nabla\zeta(\mf{x}^g).
\end{align}
Using $\mf{x}-\mf{x}^g\simeq -\zeta_\text{L}(0)\mf{x}$ around the origin and performing Fourier transformation {$\zeta^g(\mf{x}^g)\to\zeta_{\mf{k}^g}^g$}, we obtain modification of the perturbations from background modes as 
\begin{align}
    \zeta_{\mf{k}^g}^g=\zeta_{\mf{k}^g}+\zeta_\text{L}(0)\qty(3+k^g\pdif{k^g})\zeta_{\mf{k}^g}+\order{\zeta_\text{L}^2(0)}.
    \label{eq:loe_corr}
\end{align}
Let us assume that correlation functions for $\zeta_{\mf{k}^g}^g$ are observables for a local observer. 
Hereafter of this section, instead of writing ${\mf{k}^g}$, we just write $\mf{k}$ for notational simplicity.
The conjugate momenta in geodesic coordinate is discussed in~\cite{Tanaka:2012wi}.
The expression reads $\pi^g(\mf{x}^g)=e^{-3\zeta_\text{L}}\pi(\mf{x})$ in position space and
\begin{align}
    \pi^g_{\mf{k}}\simeq\pi_\mf{k}+\zeta_\text{L}(0)k\frac{\del}{\del k}\pi_\mf{k}
    \label{eq:momentum_loe}
\end{align}
in Fourier space. Also, the coordinate transformation is shown to be a canonical transformation, which preserves the commutation relation,
\begin{align}
    [\zeta^g(t,\mf{x}^g),\pi^g(t,\mf{y}^g)]\simeq ie^{-3\zeta_\text{L}}\delta^{(3)}(\mf{x}-\mf{y})\simeq i\delta^{(3)}(\mf{x}^g-\mf{y}^g).
    \label{eq:commutation}
\end{align}
{Note that the cancellation of IR contributions in loop order~\cite{Urakawa:2009my,Urakawa:2010kr,Tanaka:2012wi,Tanaka:2013xe,Tanaka:2014ina,Tanaka:2017nff} supports the commutation relation to be valid in loop level.}
As mentioned in Appendix~\ref{sec:loe}, the leading contributions of bispectra in squeezed limit is modified to less divergent behavior, showing that the interactions between long modes and short modes in superhorizon scale are turned off in the geodesic coordinate.

The IR finiteness of correlation functions {of $\zeta^g$ and $\pi^g$} was proven in~\cite{Tanaka:2017nff}.
The IR divergence indicates that the short mode is sensitive to the large scale fluctuations. 
They showed that the large-scale fluctuations work as the inhomogeneous dilatation of the metric {and confirmed that correlators of operators which are invariant under this transformation are protected against the radiative corrections from deep IR modes.} In fact, $\zeta^g$ and $\pi^g$ are invariant under the {inhomogeneous} dilatation. 

As discussed in Section~\ref{sec:eff_state}, we {may define an effective quantum state tomographically using correlation functions of $\zeta^g$ and $\pi^g$.} For example, the effective Gaussian Wigner function for the observer's system $\mf{q}$ is
\begin{align}
    W^g_\mf{q}[\zeta^g,\pi^g]\equiv \frac{1}{\det\mc{M}}\exp[-\frac{1}{2V}\Phi^{g,\alpha}(\mc{M}^{g}_{\alpha\beta})^{-1}\Phi^{g,\beta}],
\end{align}
where $\Phi^g=(\zeta^g_\mf{q},\pi^g_\mf{q})$ and $\mc{M}_{\alpha\beta}^g=\braket{\{\Phi^{g,\alpha},\Phi^{g,\beta}\}}'$, and its IR regularity is {ensured} by the IR regularity of loop corrections to power spectrum.
{Thus, the results of~\cite{Tanaka:2017nff} ensure that the effective quantum state is protected against the loop corrections from the deep IR modes.} 
This formal proof would be valid generally beyond Gaussian and one-loop order.

\subsection{IR finiteness from wavefunction approach}\label{sec:wave_ir}
{Although the previous discussion already demonstrates the IR convergence, it is instructive to perform explicit calculations using the wavefunction.}
In this subsection, we consider the transformation of the wavefunction coefficients under the coordinate transformation $\mf{x}\mapsto \mf{x}^g$.
Specifically, we will take care of the following modifications due to the coordinate transformation:
\begin{itemize}
    \item Mixing between short modes and long modes shown in \eqref{eq:loe_corr}.
    \item Redefinition of the set of observable modes $\{\mf{k}_\mc{S}\}\to\{\mf{k}^g_\mc{S}\}\simeq\{e^{-\zeta_\text{L}}\mf{k}_\mc{S}\}$.
\end{itemize}

\subsubsection{Cancellations in wavefunction coefficients}
In the wavefunction~\eqref{eq:wfu}, our expansion basis is the curvature perturbations in the comoving coordinate. Here, let us change the basis to the perturbations in the geodesic coordinate, namely,
\begin{align}
    \Psi^g[\zeta^g]
    &\equiv
    \Psi[\zeta(\zeta^g)]
    \notag\\
    &=\exp\bigg[{-}\frac{1}{2}{\int_{\mf{k}_1,\mf{k}_2}}\zeta_1^g\zeta_2^g\psi_2^g(2\pi)^3\delta^{(3)}({\mf{k}_1{+}\mf{k}_2}){-}\frac{1}{6}{\int_{\mf{k}_1,\mf{k}_2,\mf{k}_3}}\zeta_1^g\zeta_2^g\zeta_3^g\psi_3^g(2\pi)^3\delta^{(3)}({\mf{k}_1{+}\mf{k}_2{+}\mf{k}_3}){+}\cdots\bigg]
    \,,
\end{align}
where we introduced an abbreviation $\zeta_i\equiv\zeta_{\mf{k}_i}$.
Thus, the relationship between $\psi_n^g$ and $\psi_n$ can be directly read from the basis transformation {\eqref{eq:loe_corr}}. The second order terms in $\zeta$ are not modified, whereas the third order obtains correction terms from the second-order as follows:
\begin{align}
    -\frac{1}{2}\int_{\mf{k}_1,\mf{k}_2}&(2\pi)^3\delta^{(3)}(\mf{k}_1+\mf{k}_2)\zeta_1^g\zeta_2^g\psi_2^g\notag\\
    \simeq
    &-\frac{1}{2}\int_{\mf{k}_1,\mf{k}_2}(2\pi)^3\delta^{(3)}(\mf{k}_1{+}\mf{k}_2)\zeta_1\zeta_2\psi_2^g\notag\\
    &-\frac{1}{2}\int_{\mf{k}_1,\mf{k}_2,\mf{k}_3,k_1\leq k_\text{L}}(2\pi)^3\delta^{(3)}(\mf{k}_1{+}\mf{k}_2{+}\mf{k}_3)\zeta_1\zeta_2\zeta_3\qty(3-k_3\frac{\del}{\del k_3})\psi_2^g(k_2)\notag\\
    &+\order{\psi_4}
    \,,
\end{align}
{where $k_\text{L}$ denotes the superhorizon scale which is much longer than the scale of interest, i.e.,  $k_\text{L}\ll k_1, k_2$.} 
From the first term on the right-hand sides, we obtain $\psi_2^g=\psi_2$.
The second term on the right-hand sides are correction to $\psi_3$, thus the leading contribution of $\psi_3^g$ in squeezed limit is
\begin{align}
    \lim_{k_1\to 0}\psi_3^g&=\lim_{k_1\to 0}\psi_3-\qty(3-k_2\frac{\del}{\del k_2})\psi_2(k_2).
\end{align}
It exactly cancels with Ward identity~\eqref{consistency}.
Cancellation at the next leading order in $k_1$ can also be explicitly shown, as expected. 
Another note is that the same procedure shows cancellations for tensor perturbations as well, which we put in Appendix~\ref{sec:loe_tensor} for clarity of the main text.

\subsubsection{Expectation values and density matrices}
In this subsection, we confirm that $\braket{\zeta_1^g\zeta_2^g\zeta_3^g}$ is expressed by $\psi_3^g$ by considering the coordinate transformation of the integral in expectation values.
Firstly, we begin with more generic set up, i.e., observing an operator $\widehat{\mc{A}}_\text{w}^g[\widehat{\zeta}^g,\widehat{\pi}^g]$ where the index w indicates Weyl ordering. 
Also, we assume that $\pi_\text{L}$ is not included in {$\widehat{\mc{A}}_\text{w}^g$} for simplicity.
The expectation value is
\begin{align}
    \braket{\widehat{\mc{A}}_\text{w}^g[\widehat{\zeta}^g,\widehat{\pi}^g]}=\int D\zeta D\widetilde{\zeta}\braket{\widetilde{\zeta}|\widehat{\mc{A}}_\text{w}^g|\zeta}\Psi[\zeta]\Psi^*[\widetilde{\zeta}].
\end{align}
In order to define the quantum state in geodesic coordinate, we first insert complete set of $\pi$ and change the basis to
\begin{align}
    \zeta_c\equiv \frac{\zeta+\widetilde{\zeta}}{2},\qquad \zeta_\Delta\equiv\zeta-\widetilde{\zeta},
\end{align}
satisfying $D\zeta D\widetilde{\zeta}=D\zeta_c D\zeta_\Delta$, i.e.,
\begin{align}
    \braket{\widehat{\mc{A}}_\text{w}^g}=\int D\zeta_c D\zeta_\Delta D\pi \mc{A}_\text{w}^g[\zeta^g_c,\pi^g]e^{-i\int_\mf{k}\pi\zeta_\Delta}\Psi\qty[\zeta_c+\frac{\zeta_\Delta}{2}]\Psi^*\qty[\zeta_c-\frac{\zeta_\Delta}{2}].
\end{align}
Since $\pi_\text{L}$ is not included in the operator in our setup, we can perform the integration of $\pi_\text{L}$ to obtain $\delta(\zeta_{\text{L},\Delta})$. In this case, every $\zeta_\text{L}$ appearing in the expression is identical, and $e^{-i\int_\mathbf{k}\pi\zeta}=e^{-i\int_\mathbf{k}\pi^g\zeta^g}$, $D\zeta_\Delta D\pi=D\zeta_\Delta^g D\pi^g$ {are satisfied.}
These transformation properties are expected to be valid for the canonical transformation.
Then, we perform the 
transformation $\zeta\to\zeta^g$, leading to
\begin{align}
    \braket{\widehat{\mc{A}}_\text{w}^g}&=\int D\zeta_c^g D\zeta_\Delta^g D\pi^g\ \qty|\vdiff{\zeta_c}{\zeta_c^g}|\qty(\mc{A}_\text{w}^g\qty[\zeta^g_c,i\frac{\delta}{\delta\zeta_\Delta^g}]e^{-i\int_\mf{k}\pi^g\zeta_\Delta^g})\Psi^g\qty[\zeta^g_c+\frac{\zeta^g_\Delta}{2}]\Psi^{g*}\qty[\zeta^g_c-\frac{\zeta^g_\Delta}{2}]    \notag\\
    &=\int D\zeta_c^g\qty(\mc{A}_\text{w}^g\qty[\zeta^g_c,-i\frac{\delta}{\delta\zeta_\Delta^g}]\qty|\vdiff{\zeta_c}{\zeta_c^g}|\Psi^g\qty[\zeta^g_c+\frac{\zeta^g_\Delta}{2}]\Psi^{g*}\qty[\zeta^g_c-\frac{\zeta^g_\Delta}{2}])_{\zeta_\Delta^g=0},
    \label{eq:corr_loe}
\end{align}
where $|\delta\zeta_c/\delta\zeta_c^g|$ is the Jacobian.
From Eq.~\eqref{eq:corr_loe}, we obtain the quantum state
\begin{align}
    \rho_{\text{tot}}^g\qty[\zeta^g_c+\frac{\zeta^g_\Delta}{2},\zeta^g_c-\frac{\zeta^g_\Delta}{2}]\equiv \qty|\vdiff{\zeta_c}{\zeta_c^g}|\Psi^g\qty[\zeta^g_c+\frac{\zeta^g_\Delta}{2}]\Psi^{g*}\qty[\zeta^g_c-\frac{\zeta^g_\Delta}{2}].
    \label{eq:pure_g_rho}
\end{align}
{Here,} the Jacobian remains; {in fact, it gives rise to a relevant contribution to the consistency relation in the momentum bispectra discussed in Appendix~\ref{sec:loe}.}

Let us show that the bispectrum of $\zeta^g$, 
\begin{align}
    \braket{\zeta_1^g\zeta_2^g\zeta_3^g}&=\int D\zeta^g\ \qty|\vdiff{\zeta}{\zeta^g}|\ |\Psi^g[\zeta^g]|^2\zeta_1^g\zeta_2^g\zeta^g_3,
\end{align}
is not affected by the Jacobian and is expressed in terms of $\psi_3^g$ at the tree level. 
The Jacobian is calculated from the transformation~\eqref{eq:loe_corr}, \eqref{eq:momentum_loe}, and is effectively written by diagonal elements when treating $\zeta_\text{L}(0)$ as a zero mode. 
Using the perturbative expression of the Jacobian,
\begin{align}
    \qty|\vdiff{\zeta}{\zeta^g}|= 1-\int_\mathbf{k}\vdif{\zeta_{\mf{k}}^g}\qty(\zeta_\text{L}(0)\qty(3+k\pdif{k})\zeta_{\mf{k}}^g)+\order{\zeta_\text{L}^2(0)},
    \label{eq:jaco}
\end{align}
the correlation function reads
\begin{align}
    \braket{\zeta_1^g\zeta_2^g\zeta_3^g}=(2\pi)^3&\delta^{(3)}(\mf{k}_1+\mf{k}_2+\mf{k}_3) \frac{-2\Re\psi_3^g}{\prod_i 2\Re\psi_2(k_i)}\notag\\
    &+\int D\zeta^g\frac{d^3\mf{k}}{(2\pi)^3}\  \zeta_\text{L}(0)\qty(3+k\pdif{k})\zeta_{\mf{k}}^g \vdif{\zeta_{\mf{k}}^g}\qty({|\Psi[\zeta^g]|}^2\zeta_1^g\zeta_2^g\zeta^g_3)+\order{\zeta_\text{L}^2(0)}
    \label{eq:jacobian_corr}
\end{align}
after integrating by parts.
It is easy to check that the second term is subdominant when $\braket{\zeta_\text{L}\zeta_{k_i}}=0$. However, we need to consider the case $\braket{\zeta_\text{L}\zeta_{k_i}}\neq 0$ when we consider the bispectum in the squeezed limit. 
In this case, it is convenient to separate the second term in the following way:
\begin{align}
    \delta_1&\equiv \int D\zeta^g\frac{d^3\mf{k}}{(2\pi)^3}\ \zeta_\text{L}(0)\qty(3+k\pdif{k})\zeta_{\mf{k}}^g\times \zeta_1^g\zeta_2^g\zeta^g_3\vdif{\zeta_{\mf{k}}^g}\qty({|\Psi[\zeta^g]|}^2), \\
    \delta_2&\equiv \int D\zeta^g\frac{d^3\mf{k}}{(2\pi)^3}\ \zeta_\text{L}(0)\qty(3+k\pdif{k})\zeta_{\mf{k}}^g\times {|\Psi[\zeta^g]|}^2\vdif{\zeta_{\mf{k}}^g}\qty(\zeta_1^g\zeta_2^g\zeta^g_3).
\end{align}
After some perturbative calculations, we obtain 
\begin{align}
    \delta_1=-\delta_2=\lim_{k_3\to 0}\qty(3+k_1\pdif{k_1})\frac{-(2\pi)^3\delta^{(3)}(\mf{k}_1+\mf{k}_2)}{2\Re\psi_2(k_1)\Re\psi_2(k_3)}
\end{align}
for connected diagrams. It confirms that the second term in \eqref{eq:jacobian_corr} vanishes at the leading order. 
We leave complete discussions on the Jacobian for future projects.

\subsubsection{IR finiteness of the bubble loop}
{The above discussion explicitly demonstrates the wavefunction approach for the cancellation in $\braket{\zeta^g_1\zeta^g_2\zeta^g_3}$.
In Appendix~\ref{sec:loe}, we also confirm the cancellation of $\order{1/k_\text{L}^3}$ terms for momentum bispectra thanks to the consistency relation. Based on the discussion of the effective quantum state, the IR finiteness of the bubble loop is guaranteed by the behavior of the bispectra}{; to see this, let us} 
parametrize the density matrix of the full system as
\begin{align}
    \widetilde{\rho}_\text{tot}^g[\zeta^g,\widetilde{\zeta}^g;\tau]=\mc{N}\exp[\sum_{n,\ \xi=\zeta,\widetilde{\zeta}^*,\ \text{c.c.}}\qty(\prod_{j=0}^{n}\int_{\mf{k}_j}\xi^g_{\mf{k}_j}(\tau))\lambda^{\{\xi^g_{\mf{k}_i}\}}_n(\{k_i\};\tau)(2\pi)^3\delta^{(3)}\Big(\sum_i\mf{k}_i\Big)].
    \label{eq:rho_tot}
\end{align}
Note that we do not assume pure state, concerning that, for example, the Jacobian in~\eqref{eq:jaco} may include mixing terms of $\zeta$ and $\widetilde{\zeta}$.
In the same manner as the wavefunction, the three point vertices $\lambda_3^g$ correspond to tree level bispectra, and they produce bubble loop diagrams when tracing out the environment.

{The absence of dangerous $\order{1/k_\text{L}^3}$ contributions in squeezed limit of bispectra ensures that} 
$\lambda_3^g$ in squeezed configuration have more moderate behavior than the original $\psi_3$. This leads to IR finiteness of the bubble loop integral \eqref{formula} in the geodesic coordinate, where the integral exhibits only logarithmic IR divergence in the comoving coordinate.\footnote{{Decoherence may happen from mixing terms of $\zeta^g$ and $\widetilde{\zeta}^g$ for $\lambda^g_2,\lambda^g_4$ in \eqref{eq:rho_tot} as well, when it is not pure state. In this paper, we concentrate on the bubble loop contribution from $\lambda_3^g$, and the complete discussion on this point is postponed to future projects.}}
This result shows the suppression of correlations between long and short modes, {$\rho^g_\text{tot}\approx\rho_\text{long}^g\rho_\text{short}^g$}, up to tree-level cubic couplings. This is quite suggestive and is reminiscent of the factorization of the deep IR sector when evaluating correlation functions of $\zeta^g$~\cite{Tanaka:2017nff}. It would be valuable to see in future projects whether such a factorization of IR modes works in higher order as well.

Lastly, we comment on the cancellation concerned in \cite{Nelson:2016kjm}. The leading contribution for the inflationary decoherence from bulk terms is from horizon-scale environment shorter than the system modes, and they commented on the possibility of the absence of the contribution in geodesic coordinate. 
However, the cancellation actually does not occur. 
{To see this, consider the situation where the long mode $\zeta_\text{L}$ is the system mode and the environment are shorter modes. The reduced density matrix $\rho_\sys[\zeta_\text{L},\widetilde\zeta_\text{L}]$ is obtained from the wavefunction schematically as   
\begin{align}
    \rho_\sys[\zeta_\text{L},\widetilde\zeta_\text{L}]
    =
    \int D\zeta_\mc{E}
    \Psi[\zeta_\text{L},\zeta_\mc{E}]\Psi^*[\widetilde\zeta_\text{L},\zeta_\mc{E}]
    \,.\label{s_long}
\end{align}
Since the values of long mode $\zeta_\text{L}$ and $\widetilde\zeta_\text{L}$ can be different, the system-environment couplings in $\Psi[\zeta_\text{L},\zeta_\mc{E}]$ and $\Psi^*[\widetilde\zeta_\text{L},\zeta_\mc{E}]$ are not simultaneously removable via the field transformation $\zeta_\mc{E}\to \zeta^g_\mc{E}$ unless $\zeta_\text{L}=\tilde\zeta_\text{L}$. For instance, when we define $\zeta^g_\mc{E}$ using the geodesic coordinate associated with $\widetilde\zeta_\text{L}$, the system-environment couplings in $\Psi[\zeta_\text{L},\zeta_\mc{E}]$ are not removed.}
Thus, the contributions from shorter environment cannot be turned off except for diagonal components, namely $\braket{\zeta_1^g\zeta_2^g\zeta_3^g}$. 
{In fact, whatever the field transformation of environment variables is considered, it is just the change of integration measure of partial trace which does not change the final result. 

This is in contrast to the setup discussed in this section; i.e., the case where the decoherence of the system is caused by the superhorizon environment at scales much larger than that of the system. In this case, the system-environment couplings are removed after rewriting the system variables into those in the geodesic coordinate as $\zeta_\sys\to\zeta^g_\sys$. Here, the system variables are changed and hence the purity defined in terms of $\rho^g[\zeta_\sys^g,\widetilde\zeta_\sys^g]$ is also different from the one defined using $\rho[\zeta_\sys,\widetilde\zeta_\sys]$ in the comoving coordinate.}
The absorption of long mode to metric is based on the fact that the constant background mode is not an observable, and it does not make sense when two long modes, say $\zeta_\text{L}$ and $\widetilde{\zeta}_\text{L}$ in $\rho_\mc{S}[\zeta,\widetilde{\zeta}]$, are assumed to be distinguishable.

\section{Decoupling of UV physics in time-smeared observables}
\label{sec:UVdiv}

\subsection{Time-smeared observables and quantum state}
We substitute \eqref{3zetasoft} into \eqref{formula} and find that $C_q(\tau)$ is UV divergent, which naively implies the divergent decoherence rate. Presumably this divergence is due to the derivative nature of gravitational couplings; however, such divergences have been observed even in renormalizable theories in flat space, such as $\phi^3$ model in higher spacetime dimensions $D\geq5$~\cite{Balasubramanian:2011wt}. Does this divergence imply that the quantumness of $\zeta_\mf{q}$ is completely lost? Is the decoherence rate sensitive to the UV physics? If the answers were yes, it would imply the violation of the decoupling of UV physics.

As we will see below Eqs.~\eqref{F_wavefn} and \eqref{det_equal}, this divergence can be understood as arising from loop corrections to the two-point Wightman functions 
$\langle \Phi^\alpha_\mf{q}(\tau) \Phi^\beta_{\mf{q'}}(\tau') \rangle$, 
which appears only in the equal-time limit $\tau \to \tau'$ due to the violent zero-point fluctuations in the deep UV. This observation is consistent with \cite{Agon:2014uxa,Agon:2017oia,Bucciotti:2024lvb}.
Moreover, it is not canceled with the tree contributions from counterterms~\cite{Balasubramanian:2011wt} (see \cite{Burgess:2024heo} for more recent discussions). 
This is not surprising, since $\rho_\sys(\tau)$ encodes the information of equal-time correlators and is inherently sensitive to this singularity, and the divergence is found in the non-separable piece of $\rho_\sys(\tau)$ to which the local counterterm diagrams do not contribute.

However, it is impossible to measure the exactly equal-time correlator due to the finiteness of detector resolution with respect to time. What we can actually observe would be the correlation functions of $\zeta$ and $\pi$ smeared over (conformal) time,
\begin{align}
    \widehat{\overline{X}}_\mf{q}(\tau)
    \equiv 
    \int\mathrm{d}\tau'\widehat X_\mf{q}(\tau')W(\tau',\tau)
    \,,\qquad
    X_\mf{q}=\zeta_\mf{q}, \pi_\mf{q}
    \,.
\end{align}
Here, $W(\tau,\tau')$ denotes a smooth window function which is exponentially suppressed beyond the smearing scale $|\tau-\tau'|\gtrsim\delta\tau$ and is properly normalized as $\int\mathrm{d}\tau'W(\tau,\tau')=1$. The high-frequency pieces of its Fourier conjugate are then exponentially suppressed. 
A simple example is a Gaussian function $W(\tau,\tau')=(2\pi\delta\tau^2)^{-1/2}\,\exp[-\frac{(\tau-\tau')^2}{2\delta\tau^2}]$. 
In this study, we focus on the following case, 
\begin{align}
    \Lambda_\text{cutoff}^{-1}\ll\delta\tau \ll q^{-1}
    \,,\label{choice}
\end{align}
where the upper bound is imposed to ensure that we can probe the system modes $\zeta_\mf{\pm q}$ whose typical time scale is $\sim q^{-1}$, whereas the lower bound ensures that we probe only the frequency below the comoving cutoff energy scale $\Lambda_\text{cutoff}$ of the effective field theory. 
In reality, an appropriate value of $\delta \tau$ will be determined by the actual experimental setup. For instance, when considering the direct measurement of primordial gravitational waves, the resolution may simply be determined by the detector instruments. On the other hand, when measuring curvature perturbations via CMB photons,
the photons would play a role of detector apparatus. In this case, the coupling between the photons and the curvature perturbations would be important to determine $\delta\tau$, and the curvature perturbations may be effectively measured around the recombination era.
In general, the typical timescale of the coupling between the system variables of interest and the probe particles we actually measure will be important. 
While a more detailed study of this aspect would be valuable, we defer it to future work and do not specify a concrete value for $\delta\tau$ (other than the condition in \eqref{choice}) in order to preserve the generality of our present analysis.

Therefore, we propose to consider the effective quantum state defined in terms of smeared correlation functions of $\zeta$ and $\pi$, following the method developed in Section~\ref{sec:eff_state}. As we will see below, such correlators are manifestly UV finite after eliminating the standard UV divergences by renormalization. Note that the UV finiteness of time-averaged correlators is formally shown using the spectral representation in flat space in~\cite{Bucciotti:2024lvb}. 

Under the Gaussian approximation, the effective quantum state is given by the effective Wigner function $\overline W$ which is simply defined by the smeared covariance matrix $\overline\cov$,
\begin{align}
    \overline W[\zeta_\mf{q},\pi_\mf{q};\tau]
    \simeq
    \eval{\eqref{wigner_G}}_{\cov\to\overline\cov}
    \,,\,\qquad
    \overline\cov_{\alpha\beta}\equiv \langle\{\overline\Phi^\alpha_{\bf q}(\tau),\overline\Phi^\beta_{\bf q'}(\tau)\}\rangle'
    \,,
    \label{Swigner_G}
\end{align}
and hence the effective purity is determined by $\det\overline\cov$ just analogous to \eqref{purity_wigG},
\begin{align}
    \overline{P}_\sys 
    \simeq
    \frac{1}{4\det\overline\cov}
    \,.\label{Spurity_G}
\end{align}
Below, we adopt the Gaussian approximation and evaluate $\overline\cov$ up to one-loop level to obtain $\overline{P}_\sys$ via this formula \eqref{Spurity_G}.
Note that we adopt the conventional $i\epsilon$-prescription, where a positive infinitesimal parameter $\epsilon$ regulates (defines) the UV integral of the rapidly oscillating modes, and the limit $\epsilon\to 0$ is taken at the end of the calculations. See \cite{Burgess:2024heo} (and also \cite{Burgess:2024eng}) for related discussions using the $i\epsilon$-prescription.

Under the condition \eqref{choice}, the mode functions of the system variables are unaffected by the smearing procedure:
\begin{align}
    \int\mathrm{d}\tau'\,W(\tau,\tau') u_q^\alpha(\tau')
    =
    u_q^\alpha(\tau)[1+\mathcal{O}(q\delta \tau)]
    \simeq 
    u_q^\alpha(\tau)
    \,.\label{smode}
\end{align}
Therefore, the purity and the canonical commutation relation are preserved in a good approximation under the smearing procedure at the tree level. 

At loop level, however, the effects of smearing become more nontrivial since the system modes are coupled to environment modes $k\gg q$ whose typical time scale is much shorter than $q^{-1}$. This may be represented by an inequality $\zeta_\mf{q}(\tau)\neq\overline{\zeta}_{\mf{q}}(\tau)$ for the system operator $\zeta_\mf{q}(\tau)$ in the Heisenberg picture, where the time evolution is governed by the Hamiltonian which involves both slow and rapid modes.

{In this section, we do not assume that our analysis is restricted to dynamics during or after inflation. We only assume that the system modes $\pm \mf{q}$ are at superhorizon scales.}

\subsection{One-loop corrections}\label{sec:oneloop}

We compute the one-loop corrections to $\langle\Phi^\alpha_{\bf q}(\tau_1)\Phi^\beta_{\bf q'}(\tau_2)\rangle'$,
which are then used to evaluate the smeared equal-time two-point functions as follows,
\begin{align}
    \langle
    \overline\Phi^\alpha_{\bf q}(\tau)\overline\Phi^\beta_{\bf q'}(\tau)
    \rangle'
    =\int\mathrm{d}\tau_1\int\mathrm{d}\tau_2\,
    W(\tau_1,\tau)W(\tau_2,\tau)
    \langle
    \Phi^\alpha_{\bf q}(\tau_1)\Phi^\beta_{\bf q'}(\tau_2)
    \rangle'
    \,.
\end{align}
We consider only the regime $|\tau_1-\tau_2|\lesssim\delta\tau$. 
Under the condition \eqref{choice}, we ignore tiny $\mathcal{O}(q\delta\tau)$ terms and use the following replacement in the calculations below,
\begin{align}
    u_q^\alpha(\tau_{1,2})\mapsto u_q^\alpha(\tau)
    \,.\label{replace}
\end{align}
We emphasize that this replacement is done only for the mode function of the system. For convenience, we impose $\tau_1\geq\tau_2$ below, and compute $\langle\Phi^\alpha_{\bf q}(\tau_1)\Phi^\beta_{\bf q'}(\tau_2)\rangle'$ and its reverse-ordered counterpart $\langle\Phi^\beta_{\bf q}(\tau_2)\Phi^\alpha_{\bf q'}(\tau_1)\rangle'$. Note that we focus on the contributions from momenta greater than $q$ in this Section~\ref{sec:UVdiv}; however, we do not explicitly restrict the loop momenta to be smaller than $q$ in the formal expression of one-loop integrals like \eqref{F_wavefn} for simplicity.

We focus on one-loop diagrams composed of cubic vertices, 
particularly the bubble diagrams for a while. 
As is well known, the Schwinger-Keldysh formalism involves two types of fields, ``$+$'' and ``$-$'' fields, associated with a time-folded contour slightly shifted away from the real axis (represented by the dotted line) to implement the $i\epsilon$ factor:
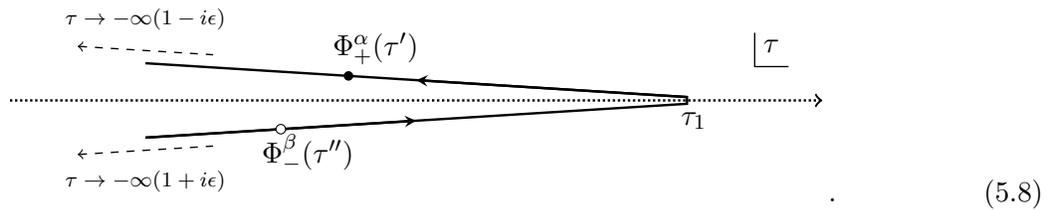
\begin{equation}
    \begin{tikzpicture}[scale=0.9]
     \draw[solid, line width=0.9pt] (-3,0.55) -- (5,0.05) -- (5,-0.05) -- (-3,-0.55);
     \draw[<-, >=stealth, line width=0.9pt] (1,0.3) -- (5,0.05);
     \draw[<-, >=stealth, line width=0.9pt] (1,-0.3) -- (-3,-0.55);
     \draw[->, densely dotted, line width=0.9pt] (-5,0) -- (7,0);
     \draw[solid, line width=0.5pt] (6,1) -- (6,0.5) -- (6.5,0.5);
     \draw (6.25,0.8) node {$\tau$};
     \draw (5.1,-0.3) node {$\tau_1$};
      \fill (0,0.3625) circle (2pt);
      \draw[fill=white] (-1,-0.425) circle (2pt);
      \draw (0.4,0.77) node {$\Phi_+^\alpha(\tau')$};
      \draw (-0.6,-0.77) node {$\Phi_-^\beta(\tau'')$};
      \draw[<-, dashed, line width=0.5pt] (-4,0.8) -- (-2,0.675);
      \draw[<-, dashed, line width=0.5pt] (-4,-0.8) -- (-2,-0.675);
      \draw (-3,1.2) node {\scriptsize $\tau\to -\infty(1-i\epsilon)$};
      \draw (-3,-1.2) node {\scriptsize $\tau\to -\infty(1+i\epsilon)$};
   \end{tikzpicture}
   \label{fig:contour}
   \,.
\end{equation}
In this figure, the fields on the upper (lower) contour are identified as $+(-)$ fields.
Accordingly, we have four types of contour-ordered propagators of $\sigma$ and $\sigma'$-fields for given $(\alpha,\beta)$,
\begin{align}
    G_{\sigma\sigma'}^{\alpha\beta}(\tau,\tau';k)
    \equiv
    \langle
        \mathcal{T}_C[\Phi^\alpha_\sigma(\tau;\mf{k})\Phi^\beta_{\sigma'}(\tau';\mf{k'})]
    \rangle'
    \,.
\end{align}
For instance, $G_{++}^{\alpha\beta}(\tau,\tau';k)$ is the Feynman propagator and $G_{+-}^{\alpha\beta}(\tau,\tau';k)$ denotes the Wightman function 
with putting the ``$-$"-field on the left, i.e., $G_{+-}^{\alpha\beta}(\tau,\tau';k)=\langle\Phi^\alpha_\mf{k}(\tau)\Phi^\beta_\mf{k'}(\tau')\rangle'$. For a review of the Schwinger-Keldysh formalism, see e.g., \cite{Kamenev:2009jj}.

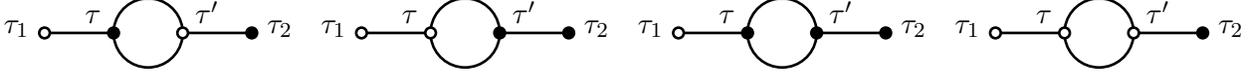
\begin{figure}[t]
\centering
\begin{fmffile}{one-loop-mix}
\begin{align*}
\begin{fmfgraph*}(98,20)
\fmfleft{i0,i1}
\fmfright{o0,o1}
\fmf{plain,tension=2}{i1,v1}
\fmf{plain,tension=2}{v2,o1}
\fmf{plain,left=1}{v1,v2,v1}
\fmfv{d.shape=circle, d.size=4}{v1,o1}
\fmfv{decor.shape=circle,decor.filled=0,decor.size=4}{v2}
\fmfv{decor.shape=circle,decor.filled=0,decor.size=4}{i1}
\fmflabel{$\tau_1$}{i1}
\fmflabel{$\tau_2$}{o1}
\fmflabel{$\tau$}{v1}
\fmflabel{$\tau'$}{v2}
\end{fmfgraph*}
\qquad
\begin{fmfgraph*}(98,20)
\fmfleft{i0,i1}
\fmfright{o0,o1}
\fmf{plain,tension=2}{i1,v1}
\fmf{plain,tension=2}{v2,o1}
\fmf{plain,left=1}{v1,v2,v1}
\fmfv{d.shape=circle, d.size=4}{v2,o1}
\fmfv{decor.shape=circle,decor.filled=0,decor.size=4}{i1,v1}
\fmflabel{$\tau_1$}{i1}
\fmflabel{$\tau_2$}{o1}
\fmflabel{$\tau$}{v1}
\fmflabel{$\tau'$}{v2}
\end{fmfgraph*}
\qquad
\begin{fmfgraph*}(98,20)
\fmfleft{i0,i1}
\fmfright{o0,o1}
\fmf{plain,tension=2}{i1,v1}
\fmf{plain,tension=2}{v2,o1}
\fmf{plain,left=1}{v1,v2,v1}
\fmfv{d.shape=circle, d.size=4}{v1,v2,o1}
\fmfv{decor.shape=circle,decor.filled=0,decor.size=4}{i1}
\fmflabel{$\tau_1$}{i1}
\fmflabel{$\tau_2$}{o1}
\fmflabel{$\tau$}{v1}
\fmflabel{$\tau'$}{v2}
\end{fmfgraph*}
\qquad
\begin{fmfgraph*}(98,20)
\fmfleft{i0,i1}
\fmfright{o0,o1}
\fmf{plain,tension=2}{i1,v1}
\fmf{plain,tension=2}{v2,o1}
\fmf{plain,left=1}{v1,v2,v1}
\fmfv{d.shape=circle, d.size=4}{o1}
\fmfv{decor.shape=circle,decor.filled=0,decor.size=4}{v1,v2,i1}
\fmflabel{$\tau_1$}{i1}
\fmflabel{$\tau_2$}{o1}
\fmflabel{$\tau$}{v1}
\fmflabel{$\tau'$}{v2}
\end{fmfgraph*}
\end{align*}
\end{fmffile}
\caption{One-loop In-In diagrams for $\langle\Phi^\alpha(\tau_1)\Phi^\beta(\tau_2)\rangle$ from cubic vertices. Vertices corresponding to plus fields are represented by black dots, while those corresponding to minus fields are represented by black circles. 
The integration domain for the vertex integral is given by $\tau\leq\tau_1$ and $\tau'\leq\tau_1$.}
\label{fig:oneloop}
\end{figure}

Depending on the assignment of the plus and minus fields, there are four possible configurations, as shown in Fig.~\ref{fig:oneloop}. Here, vertices corresponding to plus fields are represented by black dots, while those corresponding to minus fields are represented by black circles.  
We label the four contributions according to their field assignments as ``$+-$'', ``$-+$'', ``$++$'', and ``$--$'' respectively. 
The integration domain for the vertex integrals is defined by the conditions $ \tau \leq \tau_1$ and $ \tau' \leq \tau_1$. For convenience, we further divide the integration range for $\tau'$ into two regions, namely $ \tau' \leq \tau_2$ and $ \tau_2 < \tau' \leq \tau_1$. We then denote the contribution from the region $ \tau' \leq \tau_2$ by $I_{\sigma\sigma'}^{\alpha\beta}$ and the contribution from the region $ \tau_2 < \tau' \leq \tau_1$ by $\Delta I_{\sigma\sigma'}^{\alpha\beta}$, leading to   
\begin{align}
    \eval{\langle\Phi^\alpha_{\bf q}(\tau_1)\Phi^\beta_{\bf q'}(\tau_2)\rangle'}_\text{bubble}
    =
    \sum_{\sigma,\sigma'=\pm}
    \left(
    I^{\alpha\beta}_{\sigma\sigma'}(\tau_1,\tau_2)
    +
    \Delta I^{\alpha\beta}_{\sigma\sigma'}(\tau_1,\tau_2)
    \right)
    \qquad
    (\text{for $\tau_1\geq\tau_2$})
    \,.\label{oneloop1}
\end{align}
$I^{\alpha\beta}_{\sigma\sigma'}$ are evaluated as
\begin{subequations}
    \label{Idef}    
\begin{align}
    &I^{\alpha\beta}_{+-}
    =
    u_q^\alpha(\tau_1){u_q^\beta}^*(\tau_2)
    F_{+-}(\tau_1,\tau_2)
    \,,
    \qquad
    I^{\alpha\beta}_{-+}
    =
    {u_q^\alpha}^*(\tau_1){u_q^\beta}(\tau_2)
    F_{-+}(\tau_1,\tau_2)
    \,,\label{Idef_mp}
    \\
    &I^{\alpha\beta}_{++}
    =
    {u_q^\alpha}(\tau_1){u_q^\beta}(\tau_2)
    F_{++}(\tau_1,\tau_2)
    \,,
    \qquad\,\,\,
    I^{\alpha\beta}_{--}
    =
    {u_q^\alpha}^*(\tau_1){u_q^\beta}^*(\tau_2)
    F_{--}(\tau_1,\tau_2)
    \,.\label{Idef_mm}
\end{align}
\end{subequations}
Here, $F_{\sigma\sigma'}$ represent the $\alpha,\beta$-independent pieces of $I_{\sigma\sigma'}^{\alpha\beta}$. Though we record each formal expression in Appendix~\ref{sec:LoopRecord}, 
we point out that $F_{+-}$ (and its complex conjugate $F_{-+}$) is expressed in terms of the wavefunction coefficient $\psi_3$ as
\begin{align}
    F_{+-}(\tau_1,\tau_2)
    =
    \frac{1}{2}
    {u_q^{\zeta*}}(\tau_1)u_q^\zeta(\tau_2)
    \int_\mf{k}{u_k^{\zeta*}}(\tau_1)u_k^\zeta(\tau_2){u_{k'}^{\zeta*}}(\tau_1)u_{k'}^\zeta(\tau_2)
    \psi_3(q,k,k';\tau_1)\psi_3^*(q,k,k';\tau_2)
    \,,\label{F_wavefn}
\end{align}
where $k'=|\mf{k+q}|$. This is because all of propagators in the loop are of the $+-$ type which are generated only through the path integral at the final time; 
this means that this type of correction has nothing to do with quantum corrections to the bulk dynamics, in contrast to $F_{++}$ and $F_{--}$.

Moreover, $F_{+-}(\tau_1, \tau_2)$ and $F_{-+}(\tau_1, \tau_2)$ are independent of local counterterms, as the vertices arising from local counterterms in the action consist solely of either plus or minus fields. In fact, $F_{+-}$ and $F_{-+}$ are UV finite when $\tau_1\neq \tau_2$, because the integrand of \eqref{F_wavefn} oscillates very rapidly at large $k$, and its integral is defined via the $i\epsilon$-prescription as the boundary value of the convergent integral.
This ensures the UV finiteness of one-loop corrections to $\langle\Phi^\alpha_\mf{q}(\tau_1)\Phi^\beta_\mf{q}(\tau_2\neq\tau_1)\rangle'$ after renormalizing the UV divergences arising from $F_{++}$ and $F_{--}$ as usual. In fact, $F_{+-}$ and $F_{-+}$ in the current setup are UV divergent only when $\tau_1=\tau_2$ as is seen in Section~\ref{sec:div1}

\begin{figure}[t]
\centering
\begin{fmffile}{one-loop-small}
\begin{fmfgraph*}(130,40)
\fmfleft{i0,i1}
\fmfright{o0,o1}
\fmf{plain,tension=2,label=$G_\text{ret}$}{i1,v1}
\fmf{plain,tension=2,label=$G_{-+}$}{v2,o1}
\fmf{plain,left=1,label=$\frac{G_{+-}+G_{-+}}{2}$}{v1,v2}
\fmf{plain,left=1,label=$G_\text{ret}$}{v2,v1}
\fmflabel{$\tau_1$}{i1}
\fmflabel{$\tau_2$}{o1}
\fmflabel{$\tau$}{v1}
\fmflabel{$\tau'$}{v2}
\end{fmfgraph*}
\end{fmffile}
\caption{One-loop In-In diagrams for $\langle\Phi^\alpha(\tau_1)\Phi^\beta(\tau_2)\rangle$ from cubic vertices shown in Fig.~\ref{fig:oneloop} are summarized into this single configuration in the region $\tau_2<\tau'<\tau<\tau_1$. $G_\text{ret}$ denotes the retarded Green's function. We omit the indices $A$ and $B$ which label the types of field ($\zeta$ or $\pi$) on the external legs at the vertices $\tau$ and $\tau'$, respectively.}
\label{fig:small}
\end{figure}
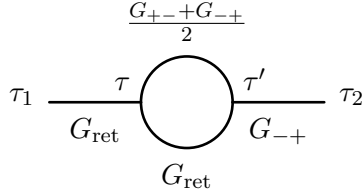

The total contributions from the integration range $\tau'>\tau_2$ are summarized into the single configuration shown in Fig.~\ref{fig:small}, 
which is evaluated as
\begin{align}
    \sum_{\sigma,\sigma'}\Delta I^{\alpha\beta}_{\sigma\sigma'}
    =
    \int^{\tau_1}_{\tau_2}\mathrm{d}\tau
    \int^{\tau}_{\tau_2}\mathrm{d}\tau'\,a^4(\tau)a^4(\tau')
    \sum_{A,B=\zeta,\pi}
    i\Im{u_q^\alpha(\tau_1){u_q^A}^*(\tau)}
    {u_q^\beta}^*(\tau_2)u_q^B(\tau') 
    iV^{AB}_{\Delta c}(\tau,\tau')
    \,.\label{deltaI}
\end{align}
Here, $V^{AB}_{\Delta c}$ denotes the amplitude of the diagram shown in Fig.~\ref{fig:small} after amputating its external lines. It is computed as the loop integral of the product of two internal propagators 
(with possible derivatives acting on them), multiplied by the appropriate coupling constants. The indices $A$ and $B$ label the types of field ($\zeta$ or $\pi$) on the external legs at the vertices $\tau$ and $\tau'$, respectively. This contribution does not exist for the case of equal-time correlators with $\tau_1=\tau_2$, since the integration range of the time integral is restricted to the domain $\tau_2<\tau'<\tau<\tau_1$.

We can also compute the two-point function with the reverse ordering, namely $\langle\Phi^\beta_{\bf q}(\tau_2)\Phi^\alpha_{\bf q'}(\tau_1)\rangle'$. The corresponding diagrams are obtained from Fig.~\ref{fig:oneloop} by swapping the plus-minus assignments for the fields at $\tau_1$ and $\tau_2$. The result is,
\begin{align}
    \langle\Phi^\beta_{\bf q}(\tau_2)\Phi^\alpha_{\bf q'}(\tau_1)\rangle'|_\text{bubble}
    =
    \sum_{\sigma,\sigma'=\pm}
    \left(
    I^{\alpha\beta}_{\sigma\sigma'}(\tau_1,\tau_2)
    +
    \Delta_2 I^{\alpha\beta}_{\sigma\sigma'}(\tau_1,\tau_2)
    \right)
    \qquad
    (\text{for $\tau_1\geq\tau_2$})
    \,,\label{oneloop2}
\end{align}
where the total contributions from the integration range $\tau'>\tau_2$ is slightly different from \eqref{deltaI}:
\begin{align}
    \sum_{\sigma,\sigma'}\Delta_2 I^{\alpha\beta}_{\sigma\sigma'}
    =
    \int^{\tau_1}_{\tau_2}\mathrm{d}\tau
    \int^{\tau}_{\tau_2}\mathrm{d}\tau'\,a^4(\tau)a^4(\tau')
    \sum_{A,B=\zeta,\pi}
    i\Im{u_q^\alpha(\tau_1){u_q^A}^*(\tau)}
    {u_q^\beta}(\tau_2){u_q^B}^*(\tau') 
    iV^{AB}_{\Delta c}(\tau,\tau')
    \,.\label{deltaI2}
\end{align}
On the other hand, the contributions from the integration range $\tau'\leq \tau_2$ are identical to those found in \eqref{oneloop1}.

In fact, the contributions from the integration region $\tau'>\tau_2 $ result in only tiny corrections to both the smeared commutation relation and the smeared purity. {For the smeared purity, the amplitude of the diagram shown in Fig.~\ref{fig:small} is much smaller than the amplitude of each diagram in Fig.~\ref{fig:oneloop}, because the retarded Green's function at superhorizon scales is much smaller than the Wightman function. This reflects the enhancement of fluctuations at superhorizon scales. Therefore,  Fig.~\ref{fig:small} gives negligible contributions for the smeared purity. For the smeared commutation relation, we can estimate the contribution from the diagram in Fig.~\ref{fig:small} by using the technique explained in Appendix~\ref{sec:nogrow} and show its smallness under the condition \eqref{choice}.}
Hence, this only gives perturbatively tiny corrections to the smeared purity and commutation relation. 
The same reasoning applies to the contributions from the diagrams shown in Fig.~\ref{fig:quartic}. Therefore, we will neglect such minor contributions; however, one can compute them from \eqref{deltaI} and \eqref{deltaI2}.\footnote{In theories without derivative interactions, these contributions are $\mathcal{O}(q\delta\tau)$-suppressed as $G_\text{ret}(\tau_1,\tau;q)$ does when $|\tau_1-\tau|\lesssim \delta\tau$. In theories with derivative interactions as in the present case, there are terms not suppressed by $\mathcal{O}(q\delta\tau)$. However, the size of loop correction to the smeared purity and commutation relation is of the order of tiny coupling constants multiplied by some inverse powers of $(\Lambda_\text{cutoff}\delta\tau)$, and the latter cannot exceed unity under the condition \eqref{choice}. Here, we suppose that the coupling strength of derivative interactions are suppressed by the EFT cutoff energy scale $\Lambda_\text{cutoff}$.}

\subsubsection{ Smeared commutation relation}
Let us now study one-loop corrections to the smeared commutation relation, 
\begin{align}
    \left\langle
    \left[
    \overline\zeta_{\bf q}(\tau),\overline\pi_{\bf q'}(\tau)
    \right]
    \right\rangle'
    =\int\mathrm{d}\tau_1\int^{\tau_1}\mathrm{d}\tau_2\,
    W(\tau_1,\tau)W(\tau_2,\tau)
    \left\{
    \left\langle
    \left[
    \zeta_{\bf q}(\tau_1),\pi_{\bf q'}(\tau_2)
    \right]
    \right\rangle'
    -
    \left\langle
    \left[
    \pi_{\bf q'}(\tau_1),\zeta_{\bf q}(\tau_2)
    \right]
    \right\rangle'
    \right\}
    \,.
\end{align}
When focusing on the contribution from the integration range $\tau' \leq \tau_2$, we find from Eqs.~\eqref{oneloop1} and \eqref{oneloop2} that the one-loop correction from bubble diagrams to the canonical commutation relation vanishes. Moreover, by straightforward calculations, we can confirm that one-loop contributions other than the bubble type shown in Fig.~\ref{fig:quartic} do not modify the commutation relation either.
Therefore, we conclude
\begin{align}
    &\eval{\left\langle
    \left[
    \zeta_{\bf q}(\tau_1),\pi_{\bf q'}(\tau_2)
    \right]
    \right\rangle'}_\text{one-loop}
    \simeq
    0
    \simeq
    \eval{\left\langle
    \left[
    \pi_{\bf q'}(\tau_1),\zeta_{\bf q}(\tau_2)
    \right]
    \right\rangle}_\text{one-loop}
    \quad
    (|\tau_1-\tau_2|\lesssim \delta\tau\ll q^{-1})
    \,,\label{com0-1}
\end{align}
where we used the symbol ``$\simeq$'' since the tiny contributions from the integration range $\tau_2<\tau'\leq\tau_1$ are ignored.
In the equal time limit $\tau_2\to\tau_1$, such terms are absent and \eqref{com0-1} shows the cancellation of one-loop corrections to the commutation relation.

\begin{figure}[t]
\centering
\begin{fmffile}{one-loop-quartic}
\begin{align*}
\begin{fmfgraph*}(100,40)
\fmfleft{i0,i1}
\fmfright{o0,o1}
\fmf{plain,tension=2}{i1,v1}
\fmf{plain,tension=2}{v1,o1}
\fmf{plain,left=1}{v1,v2,v1}
\fmf{phantom,tension=3}{i0,v2}
\fmf{phantom,tension=3}{v2,o0}
\fmflabel{$\tau_1$}{i1}
\fmflabel{$\tau_2$}{o1}
\end{fmfgraph*}
\qquad
\begin{fmfgraph*}(100,40)
\fmfleft{i0,i1}
\fmfright{o0,o1}
\fmf{plain,tension=2}{i1,v1}
\fmf{plain,tension=2}{v1,o1}
\fmf{plain,left=1}{v1,v2,v1}
\fmf{phantom,tension=3}{i0,v2}
\fmf{phantom,tension=3}{v2,o0}
\fmflabel{$\tau_1$}{i1}
\fmflabel{$\tau_2$}{o1}
\end{fmfgraph*}
\qquad
\begin{fmfgraph*}(100,40)
\fmfleft{i0,i1}
\fmfright{o0,o1}
\fmf{plain,tension=1.3}{i1,v1}
\fmf{plain,tension=1.3}{v1,o1}
\fmf{plain,tension=2.6}{v1,v2}
\fmf{plain,left=1}{v2,v3,v2}
\fmf{phantom,tension=2.7}{i0,v3}
\fmf{phantom,tension=2.7}{v3,o0}
\fmflabel{$\tau_1$}{i1}
\fmflabel{$\tau_2$}{o1}
\end{fmfgraph*}
\end{align*}
\end{fmffile}
\caption{One-loop In-In diagrams for $\langle\Phi^\alpha(\tau_1)\Phi^\alpha(\tau_2)\rangle$ other than the bubble type. 
The contributions of these diagrams to the smeared purity are negligible.}
\label{fig:quartic}
\end{figure}

\refstepcounter{subsubsection}
\subsubsection*{\thesubsubsection\quad Smeared purity}
The smeared purity is 
\begin{align}
    \overline{\mathcal{P}}_\sys(\tau)
    \simeq
    \frac{1}{1+\overline{\Delta}_q(\tau)}
    \,,\qquad
    \overline{\Delta}_q(\tau)
    =
    4\delta\det\overline\cov_\text{one-loop}
    \,,\label{purity_det}
\end{align}
where $\delta\det\overline\cov_\text{one-loop}$ is a one-loop correction to $\det\overline\cov$. Writing $\overline\cov=\overline\cov_\text{tree}+\overline\cov_\text{one-loop}$, we evaluate it ignoring higher-order $\mathcal{O}((\overline\cov_\text{one-loop})^2)$ terms.
Since the smeared covariance matrix is
\begin{align}
    \overline\cov_{\alpha\beta}
    =
    \int\mathrm{d}\tau_1\int^{\tau_1}\mathrm{d}\tau_2\,
    W(\tau_1,\tau)W(\tau_2,\tau)
    \left\langle
    \left\{\Phi^\alpha_{\bf q}(\tau_1),\Phi^\beta_{\bf q'}(\tau_2)\right\}
    +
    (\alpha\leftrightarrow \beta)
    \right\rangle'
    \,,
\end{align}
$\cov_\text{one-loop}$ is evaluated from Eqs.~\eqref{oneloop1}, \eqref{Idef}, and \eqref{oneloop2} as  
\begin{align}
    \overline\cov_{\alpha\beta,\text{one-loop}}
    \simeq
    &2\left(
    u^\alpha_q(\tau){u_q^\beta}^*(\tau)
    +
    u^\beta_q(\tau){u_q^\alpha}^*(\tau)
    \right)
    \int\mathrm{d}\tau_1\int^{\tau_1}\mathrm{d}\tau_2\,
    W(\tau_1,\tau)W(\tau_2,\tau)
    \Re{F_{+-}(\tau_1,\tau_2)}
    \no\\
    &+\mathcal{O}(u_q^\alpha(\tau) u_q^\beta(\tau),(u_q^\alpha (\tau)u_q^\beta(\tau))^*)
    \,.
    \label{Scov_loop}
\end{align}
Here, we also used the condition $F_{-+}(\tau_1,\tau_2)=F_{+-}^*(\tau_1,\tau_2)$ 
and the approximation \eqref{replace}, and ignored the tiny contributions from the integration range $\tau_2<\tau'<\tau_1$. The last term denotes the terms proportional to $u_q^\alpha(\tau) u_q^\beta(\tau)$ or $(u_q^\alpha (\tau)u_q^\beta(\tau))^*$. Such terms do not contribute to $\det\overline\cov$ at $\mathcal{O}(\overline\cov_\text{one-loop})$. From Eq.~\eqref{Scov_loop}, we evaluate the one-loop correction to $\det\overline\cov$ as
\begin{align}
    \delta\det\overline\cov_\text{one-loop}(\tau)
    &\simeq
    2
    \int\mathrm{d}\tau_1\int^{\tau_1}\mathrm{d}\tau_2\,
    W(\tau_1,\tau)W(\tau_2,\tau)
    \Re{F_{+-}(\tau_1,\tau_2)}
    \,.\label{det_loop1}
\end{align}
Substituting \eqref{F_wavefn} into \eqref{det_loop1} and using the approximation \eqref{replace} again, we obtain
\begin{align}
    \overline{\Delta}_q(\tau)
    &\simeq
    2
    P_q(\tau)
    \int_\mf{k}
    \int\mathrm{d}\tau_1\int\mathrm{d}\tau_2\,
    W(\tau_1,\tau)W(\tau_2,\tau)
    \no\\
    &\quad\times
    \Re{
    {u_k^\zeta}^*(\tau_1){u_k}^\zeta(\tau_2){u_{k'}^\zeta}^*(\tau_1)u_{k'}^\zeta(\tau_2)
    \psi_3(q,k,k';\tau_1)\psi_3^*(q,k,k';\tau_2)
    }
    \,.\label{det_loop2}
\end{align}
Here, we exchanged the order of integrals.\footnote{This is allowed since the $i\epsilon$-prescription ensures the convergence of the loop integral $\int_\mf{k}$.} Also, we eliminated the upper end of $\int\mathrm{d}\tau_2$ since the integrand is invariant under the permutation $\tau_1\leftrightarrow\tau_2$. 

There are two contributions in \eqref{det_loop2}: those from the modes satisfying 
$k > k_*(\tau)$ and those with $k < k_*(\tau)$, where $k_*(\tau)=|\tau|$ during inflation ($\tau<\tau_\text{e}$) and $k_*(\tau)=|\tau_\text{e}|$ after the end of inflation ($\tau>\tau_\text{e}$). The former did not cross the horizon during inflation and remain in the vacuum state during and after inflation, whereas the latter crossed the horizon during inflation and the fluctuations are amplified. From the late time perspective $\tau>\tau_\text{e}$, the latter can be understood as an excited environment. These two contributions are qualitatively different and are treated separately in Sections~\ref{sec:false} and \ref{sec:genuine}, respectively.

Note that we can recover the result in the equal time limit by setting $W(\tau',\tau)=\delta(\tau'-\tau)$ in \eqref{det_loop1} or \eqref{det_loop2}, leading to
\begin{align}
    \Delta_q(\tau)
    =
    4
    \Re{F_{+-}(\tau,\tau)}
    =
    2
    P_q(\tau)
    \int_\mf{k} P_k(\tau)P_{|\mf{k{+}q}|}(\tau)
    |\psi_3(q,k,|\mf{k{+}q}|;\tau)|^2
    \,,
    \label{det_equal}
\end{align}
which exactly reproduces the previous result for the perturbative purity \eqref{purity_earl} with \eqref{formula}. 
As is already seen in Section~\ref{sec:div1}, this integral is UV divergent. Eq.~\eqref{det_equal} identifies the divergence we found in the purity as the divergent one-loop correction to $\langle\Phi^\alpha_\mf{q}(\tau_1)\Phi^\beta_\mf{q}(\tau_2)\rangle'$ in the equal-time limit $\tau_1\to \tau_2$. This divergence exists only in the equal-time limit $\tau_1\to\tau_2$, since the integrand of \eqref{F_wavefn} oscillates very rapidly at large $k$ away from the equal-time limit, making the integrand convergent (see also the discussions below \eqref{F_wavefn}).
This observation is consistent with the findings of Refs.~\cite{Agon:2014uxa,Agon:2017oia}, which clarified the connection between the UV divergences in the non-Hamiltonian term of the master equation for the reduced density matrix and the renormalization of composite operators.

\subsection{False decoherence: decoupling of UV physics}\label{sec:false}
In this subsection, we focus on the contributions from the modes satisfying $k>k_*(\tau)$ which are in the vacuum state after inflation.
It is convenient to extract the oscillatory pieces of the integrand by writing the mode function as $u_k^\zeta(\tau)=A_k(\tau)e^{-ik\tau}$,
\begin{align}
    \eval{\overline\Delta_q(\tau)}_{k>k_*(\tau)}
    &\simeq
    2P_q(\tau)
    \int_{\mf{k}}
    \int\mathrm{d}\tau_1\int\mathrm{d}\tau_2\,
    W(\tau_1,\tau)W(\tau_2,\tau)
    \Theta\left(k-k_*(\tau)\right)
    \no\\
    &\quad\times
    \Re{A_k^*(\tau_1)A_{k'}^*(\tau_1)A_k(\tau_2)A_{k'}(\tau_2)
    \psi_3(q,k,k';\tau_1)\psi_3^*(q,k,k';\tau_2)
    e^{i(k+k')(\tau_1-\tau_2)}
    }
    \,,\label{det_loop3}
\end{align}
where $A_k$ and $\psi_3$ do not oscillate in time. Here, $k'=|\mf{k{+}q}|$. Hence, the integrand of Eq.~\eqref{det_loop3} oscillates rapidly at large $k$ when $\tau_1\neq \tau_2$. In particular, the contributions from subhorizon modes which also satisfy $k\geq M\sim (\delta\tau)^{-1}$ are exponentially suppressed after being integrated against the smearing function $W$; a basic mechanism for this is simply explained by the following equality,
\begin{align}
    &\int_M \d{k}
    \int\mathrm{d}\tau_1\int\mathrm{d}\tau_2\,
    W(\tau_1,\tau)W(\tau_2,\tau) 
    e^{i(k+k')(\tau_1-\tau_2)}
    =
    \int_M \d{k}\,
    e^{-(k+k')^2(\delta\tau)^2}
    \simeq
    \frac{e^{-4M^2(\delta\tau)^2}}{8 M \delta\tau}
    \,,
    \label{expsup}
\end{align}
which decays exponentially at large $M\gtrsim1/\delta\tau$, though the original integral $\int_M\mathrm{d}k e^{i(k+k')(\tau_1-\tau_2)}$ before the smearing grows linearly in $M$ in the limit $M\to\infty$! Here, we chose a Gaussian window function $W(\tau,\tau')=(2\pi\delta\tau^2)^{-1/2}\,\exp[-\frac{(\tau-\tau')^2}{2\delta\tau^2}]$ for illustration. We also performed a time integral from $-\infty$ to $\infty$, though the integration range should be finite in a realistic measurement. However, these are just technical simplifications, and the exponential suppression of UV contributions will be generic as long as the high-frequency component $\omega\gtrsim 1/\delta\tau$ of the window function is exponentially suppressed.

Thus, time smearing provides a robust mechanism for eliminating the spurious UV-divergent quantum decoherence of the system caused by the violent zero-point fluctuations of the deep UV modes satisfying $k\gtrsim \delta\tau^{-1}$. 
Eq.~\eqref{expsup} would also explain the decoupling of heavy particles {in vacuum state whose mode functions oscillate rapidly with frequency above $1/\delta\tau$ and contain positive frequency modes alone.} 
The violation of unitarity due to the ignorance of heavy physics {beyond the EFT resolution (in both space and time)} is exponentially suppressed for an observer who only measures time-averaged observables, as long as the excitations of heavy particles are negligibly tiny. {This conclusion should hold true even in flat space.}

Our result can be understood as follows.
The smearing effectively projects out the high-frequency modes involved in the system variables $\zeta_{\mf{\pm q}}(\tau)$ and $\pi_{\mf{\pm q}}(\tau)$ that are local in time. Since the slow modes cannot excite rapid modes in the environment, the rapidly oscillating environment in the vacuum state cannot spoil the quantum coherence of slowly varying components of system variables. 

Our finding is consistent with the statement of ``false decoherence'' proposed in \cite{unruh:2011fal} (see also \cite{Gundhi:2023vjs}), which claimed that the decoherence caused by the rapid environment whose time scale is much faster than that of the system is an artifact. 
An important lesson here is that {\it large entanglement does not necessarily imply a loss of coherence of the system.}

\subsubsection{Related discussions}\label{sec:related}

It would be worth mentioning some related discussions. 
As explained around \eqref{det_equal}, the UV divergence in purity is understood as the divergence of two-point correlators in the equal-time limit. This is consistent with \cite{Agon:2014uxa,Agon:2017oia}. 
Interestingly, Ref.~\cite{Burgess:2024heo} pointed out a possible role of composite operator mixing in removing UV divergences of purity. It would be valuable to pursue this direction in a way applicable to the present setup. For instance, the regularized density matrix or Wigner function may be defined such that they correctly reproduce the equal-time correlators defined properly as renormalized composite operators which make sense within EFT. It would be then interesting to clarify the relation to the approach adopted here. We leave these aspects for future work. 

The discussion in Section~\ref{sec:UVdiv} is closely related to the presence of nontrivial contributions from heavy particles to equal-time correlators, which cannot be captured by the unitary EFT defined by the conventional effective action $S_\text{EFT}$, i.e., the effective action obtained after treating heavy modes as non-dynamical degrees of freedom and integrating them out~\cite{Green:2024cmx}. The presence of such terms will be due to the high-frequency components contained in EFT fields which are local in time. Indeed, this type of contribution from UV physics will be absent in the S-matrix (and hence also in the residue of the total energy pole of wavefunction coefficients), since the on-shell condition ensures that the frequency must also be restricted within the EFT regime. 

Moreover, for the case of S-matrix, the Hilbert space is spanned by the asymptotic on-shell states in which slow and fast modes are cleanly separated. Hence, it is straightforward to trace out the fast modes. However, in general cases, it is nontrivial to single out the slow modes at the level of Hilbert space. One of the advantage of our approach is that we can avoid this issue by focusing on the time-smeared observables and considering corresponding effective quantum state.

\subsection{Genuine decoherence: excited environment}\label{sec:genuine}

From the discussion in the previous subsection, we conclude that the slowly varying pieces of the integrand of \eqref{det_loop2} whose typical frequencies measured in the conformal time are less than $\delta\tau^{-1}$ are the genuine source of decoherence, 
\begin{align}
   \overline\Delta_q(\tau)
    \approx
    2
    P_q(\tau)
    \int_\mf{k}
    \mathcal{I}_\text{slow}(k,|\mf{k{+}q}|,q;\tau,\tau)
    \,.\label{slow}
\end{align}
Here, $\mathcal{I}$ denotes the integrand of \eqref{det_loop2}, 
\begin{align}
    \mathcal{I}(k,k',q;\tau_1,\tau_2)
    \equiv 
    \Re{
    {u_k^\zeta}^*(\tau_1){u_k}^\zeta(\tau_2){u_{k'}^\zeta}^*(\tau_1)u_{k'}^\zeta(\tau_2)
    \psi_3(q,k,k';\tau_1)\psi_3^*(q,k,k';\tau_2)
    }
    \,,\label{defI}
\end{align}
and $\mathcal{I}_\text{slow}(k,k',q;\tau,\tau)$ denotes the slowly varying piece of $\mathcal{I}(k,k',q;\tau_1,\tau_2)$, which remains unchanged approximately under an arbitrary shift of $\tau_1$ and $\tau_2$ within the range $[\tau-\delta\tau,\tau+\delta\tau]$.

For the case with $\delta\tau\lesssim 1/k_*(\tau)$, only the modes satisfying $k\delta\tau\lesssim 1$ can contribute to $\mathcal{I}_\text{slow}$ as we studied in the previous subsection. In another case $\delta\tau\gtrsim 1/k_*(\tau)$, there is a nontrivial contribution to $\mathcal{I}_\text{slow}$ 
from the modes satisfying $\delta\tau^{-1}\lesssim k \lesssim k_*(\tau)$, which exit the horizon during inflation. 
The fluctuations of such modes are amplified and frozen after the horizon crossing, and they are in excited states in the late universe; i.e., their mode functions in the late universe are given by the sum of positive and negative frequency modes, schematically given by $u_k(\tau)\sim \alpha_k e^{-ik\tau}+\beta_k e^{ik\tau}$ with non-negligible $\beta_k$ (see e.g.,~\cite{Maggiore:1999vm} for more details).
Since this mixing leads to the phase cancellation, even the large-$k$ modes satisfying $\delta\tau^{-1}\lesssim k \lesssim k_*(\tau)$ contribute to $\mathcal{I}_\text{slow}$. 
Now we recast \eqref{slow} as
\begin{align}
   \overline\Delta_q(\tau)
    \approx
    2
    P_q(\tau)
    \int_\mf{k}
    \Bigl[
    &P_k(\tau)P_{k'}(\tau)
    |\psi_3(q,k,k';\tau)|^2
    \Theta\left(1-k\delta\tau\right)
    \no\\
    &
    +
    \mathcal{I}_\text{slow}(k,k',q;\tau,\tau)
    \Theta\left(k\delta\tau-1\right)
    \Theta\left(k_*(\tau)-k\right)
    \Bigr]
    \,,\label{det_oneloop4}
\end{align}
with $k'=|\mf{k{+}q}|$.
The term in the second line represents the nontrivial contributions from modes in the window $\delta\tau^{-1}\lesssim k \lesssim k_*(\tau)$. This formula clarifies the main contributions to the smeared purity and captures the main results of Section.~\ref{sec:UVdiv}. The false decoherence caused by zero-point fluctuations at deep UV modes with $k>\max[\delta\tau^{-1},k_*(\tau)]$ is absent. We conclude that the modes which cross the horizon during inflation would serve as a robust and genuine source of decoherence.\footnote{
Note that the formula \eqref{det_oneloop4} clarifies the main contributions to the smeared purity, while it involves some minor errors in the contributions from the modes $k\sim\delta\tau^{-1}$. This is because we modeled the exponential decay at $k\delta\tau\gtrsim 1$ by the step function, and we also ignored the time dependence of $u_k$ and $\psi_3$ to obtain \eqref{det_oneloop4} from \eqref{det_loop2}. A more precise formula is \eqref{det_loop2}.}

\subsubsection{An example of decoherence from excited environment}\label{sec:eg}
It may be useful to clarify the role of excited environment using Bogoliubov coefficients as a genuine source of decoherence in a simple example. Consider a $g\dot\phi^3$ theory in four-dimensional Minkowski spacetime for illustration, where $\phi$ is a scalar field with mass $m$. As we did for the case of $\zeta$, we consider a pair of modes $\pm\mf{q}$ as system modes and treat other modes as environment.
Suppose that the whole system is initially in a vacuum pure state while it evolves into an excited state at a late time due to some event which induces the time dependence of the mass $m$ at the intermediate time $\sim \tau_*$.\footnote{Such a mechanism may be provided, for example, through a coupling with time-dependent external classical sources such as homogeneous classical scalar field. Note that this setup is needed to ignore the lower-end of the time integral when we evaluate the cubic wavefunction coefficient $\psi_3$ in this model using the $i\epsilon$-prescription. For the case of $\zeta$, for instance, $\tau_*$ corresponds to the reheating epoch.} 
The mode function of $\phi$ at $\tau\gtrsim \tau_*$ is given using the Bogoliubov coefficients $(\alpha_k,\beta_k)$ as
\begin{align}
    u_k(\tau) 
    =\frac{1}{\sqrt{2\omega_k}}
    \left(
    \alpha_k e^{-i\omega_k\tau} + \beta_k e^{i\omega_k\tau}
    \right)
    \,,
\end{align}
where $\omega_k\equiv \sqrt{k^2+m^2}$ and the normalization condition $|\alpha_k|^2-|\beta_k|^2=1$ is imposed. 

In this setup, we compute the perturbative smeared purity $\mathcal{P}_\sys|_\text{$g\dot\phi^3$}\simeq 1-\overline\Delta_q|_\text{$g\dot\phi^3$}$ for the system $\phi_{\pm \mf{q}}$ at one-loop level. We choose the time averaging scale $\delta \tau$ to be $q\delta \tau\ll 1$ similarly to \eqref{choice}.
We can use the same formula as \eqref{det_loop2} but with the cubic wavefunction coefficient and the mode function given in this setup. The result is,
\begin{align}
   \eval{\overline\Delta_q(\tau)}_\text{$g\dot\phi^3$}
    &\simeq
    2
    |u_q(\tau)|^2
    \int_\mf{k}
    \int\mathrm{d}\tau_1\int\mathrm{d}\tau_2 W(\tau_1,\tau) W(\tau_2,\tau)\,\mathcal{I}_\text{$g\dot\phi^3$}(k,k',q;\tau_1,\tau_2)
    \,,\label{purity_bogo}
\end{align}
where the integrand $\mathcal I_\text{$g\dot\phi^3$}(\tau_1,\tau_2;k,k',q)$ involves the slowly varying term even at large $k\gg 1/
\delta \tau$ (we also impose $k\gg m$ for simplicity) when environments are excited $\beta_k\neq0$:
\begin{align}
    \eval{\mathcal{I}_\text{$g\dot\phi^3$, slow}(k,k',q;\tau,\tau)}_{k\gg 1/\delta \tau,m}
    \approx
    g^2k^2q^2\tau^2(|\beta_k|^2+1)|\beta_k|^2
    \,,\label{excited}
\end{align}
where we ignored the terms which are sub-dominant at a late time $\tau\gg \tau_*$.
The growth in $\tau$ implies that the information of the system is continuously recorded by the environment. The point is that the decoherence rate is proportional to $(1+|\beta_k|^2)|\beta_k|^2$, where $|\beta_k|^2$ is the number density of excited particles. Hence, no decoherence is caused by the hard modes $k\gg \delta \tau^{-1}$ in a vacuum state. This result is analogous to what happens in the case of $\zeta$.

\section{Finite decoherence rate}\label{sec:finite_deco}
Given discussions in Sections~\ref{sec:IRdiv} and \ref{sec:UVdiv}, we propose to consider time-averaged $\zeta^g$ and $\pi^g$ as observables, which are denoted by $(\overline{\zeta}^g,\overline{\pi}^g)$. We use the time-averaged $\zeta_\text{L}(0)$ to define the geodesic coordinate here to project out the high-frequency components in $\zeta_\text{L}(0)$. At the leading order in $\overline{\zeta}_\text{L}(0)$, these fields are given by
\begin{align}
    &\overline{\zeta}^g_\mf{k}
    =\overline{\zeta}_\mf{k}
    +
    \overline{\zeta}_\text{L}(0)
    \left(3+k\frac{\der}{\der k}\right)\overline{\zeta}_\mf{k}
    +\mathcal{O}\left({\overline{\zeta}^2_\text{L}(0)}\right)
    \,,\\
    &\overline{\pi}^g_\mf{k}
    =\overline{\pi}_\mf{k}
    +
    \overline{\zeta}_\text{L}(0)
    k\frac{\der}{\der k}\overline{\pi}_\mf{k}
    +\mathcal{O}\left(\overline{\zeta}^2_\text{L}(0)\right)
    \,.
\end{align}
The effective purity defined based on the correlators of $(\overline{\zeta}^g,\overline{\pi}^g)$ do not suffer from UV divergences, since the high-frequency components in them are projected out. On the other hand, it is also IR finite; the time-averaging procedure does not change the discussion of IR regularity in Section~\ref{sec:IRdiv}, since the integrand of IR loops does not oscillate in time. 

Therefore, writing the effective purity defined in this way as $\overline{\mathcal{P}}_\sys^g(\tau)=(1+\overline{\Delta}^g_q(\tau))^{-1}$,
we can calculate $\overline{\Delta}^g_q(\tau)$ at one-loop level {using Eq.~\eqref{det_oneloop4}, with contributions from deep IR environment below the system wavenumber $q$ projected out,}
\begin{align}
   \overline\Delta^g_q(\tau)
    \approx
    2
    P_q(\tau)
    \int_\mf{k}
    \Theta\left(\min[k,k']-q\right)
    \Bigl[
    &P_k(\tau)P_{k'}(\tau)
    |\psi_3(q,k,k';\tau)|^2
    \Theta\left(1-k\delta\tau\right)
    \no\\
    &
    +
    \mathcal{I}_\text{slow}(k,k',q;\tau,\tau)
    \Theta\left(k\delta\tau-1\right)
    \Theta\left(k_*(\tau)-k\right)
    \Bigr]
    \,.
    \label{IRUV_purity}
\end{align}
Here, $k'=|\mf{k{+}q}|$. This is the main formula of this study, which is manifestly IR and UV finite and can be used for the study of decoherence of $\zeta$. This formula reflects that decoherence caused by deep IR or UV modes is false and is not seen by a local observer who measures the correlation functions of $\overline\zeta^g$ and $\overline\pi^g$. This formula will be valid at least before the system modes $\pm \mf{q}$ re-enter the horizon, since our analysis assumes that the system modes are at superhorizon scales. 


In this section, we assume that the smearing scale is shorter than the Hubble time at the evaluation time $\tau$, i.e., $\delta\tau\lesssim (aH)^{-1}$, and focus on the decoherence induced by the superhorizon environment for which the usual perturbation theory is well under control:
\begin{align}
   \overline\Delta^g_q(\tau)|_{k<aH}
    \approx
    2
    P_q(\tau)
    \int_\mf{k}
    \Theta\left(\min[k,k']-q\right)
    \Theta\left(a(\tau)H(\tau)-k\right)
    &P_k(\tau)P_{k'}(\tau)
    |\psi_3(q,k,k';\tau)|^2
    \,.\label{superpurity}
\end{align}
This places a lower bound on $\overline\Delta^g_q(\tau)$. With this choice of $\delta\tau$, we evaluate $\overline\Delta^g_q(\tau)$ at any time $\tau$ after the system modes $\pm \mf{q}$ exit the horizon during inflation until they re-enter at $\tau \sim 1/q$, consistently with the condition \eqref{choice}. Note that even when considering a larger $\delta\tau$, the decoherence rate obtained in this section will not be significantly reduced. This is because the modes that crossed the horizon during inflation act as an excited environment, inducing genuine decoherence that is robust against time smearing, as discussed in Section~\ref{sec:genuine}.

\subsection{Evolution during inflation}\label{sec:before}
To begin with, we study the decoherence rate $\overline\Delta^g_q(\tau)|_{k<aH}$ caused by the superhorizon environment during inflation. 
We use the consistency relation \eqref{consistency} to evaluate $\psi_3$ from $\psi_2$. For clarity, we separately treat the contributions from $\psi_\text{2,bd}$ and $\psi_\text{2,bulk}$. The former and the latter denote the contributions from the quadratic boundary term and the bulk term in the action, respectively; their expressions are given in \eqref{eq:psi2}. 
The contribution from the former is evaluated as\footnote{
Strictly speaking, the integral of \eqref{bd_inf} is dominated by the lower end $k\sim q$. In this case, the use of consistency relation \eqref{eq:psi2} is not justified in general. However, the quadratic boundary contributions $\Im\psi_\text{2,bd}= 18\Mpl^2Ha^3-6\Mpl^2ak^2/H$ and \eqref{eq:psi2} correctly predict the terms $54\Mpl^2Ha^3-2\Mpl^2ak^2/H$ of $\im\,\psi_3$ even when $k\sim q$, as long as $q$ is at superhorizon scales. The corresponding interaction terms are in the cubic boundary Lagrangian $\Lag_\text{int}\ni \der_t[-9a^3H\zeta^3+\frac{a}{H}\zeta(\der_i\zeta)^2]$. Hence, the use of consistency relation in \eqref{bd_inf} gives a correct result.}
\begin{align}
    \eval{\overline{\Delta}^g_q(\tau)}_{k<aH,\psi_\text{2,bd}}
    &\approx
    2
    P_\zeta(q;\tau)
    \int^{k<-\frac{1}{\tau}}_{k,k'>q}\frac{\mathrm{d}^3k}{(2\pi)^3} 
    P_\zeta(k;\tau)P_\zeta(k';\tau)
    \left|
    \left(3-k\frac{\der}{\der k}\right)
    \Im\psi_\text{2,bd}(k;\tau)
    \right|^2
    \no\\
    &\approx
    \frac{729(16 \ln2-9) H^2}{256 \pi ^2  \Mpl^2\epsilon ^3}\frac{1}{(-q\tau)^6}
    \,.\label{bd_inf}
\end{align}
The bulk contribution is 
\begin{align}
    \eval{\overline{\Delta}^g_q(\tau)}_{k<aH,\psi_\text{2,bulk}}
    &\approx
    2P_\zeta(q;\tau)
    \int_q^{-1/\tau}\frac{\mathrm{d}k}{2\pi^2}k^2 
    P_\zeta^2(k;\tau)
    \left|
    \left(3-k\frac{\der}{\der k}\right)
    \psi_\text{2,bulk}(k;\tau)
    \right|^2
    \no\\
    &\approx
    \frac{H^2}{16\pi ^2 \Mpl^2\epsilon}
    \frac{1}{(-q\tau)^3}
    \left[
        \frac{(1-n_s)^2}{3} + \frac{5(4-\pi)}{2}
    \right]
    \,.\label{bulk_inf}
\end{align}
Here, the integral is dominated by the upper end $k\sim -1/\tau$, and hence the use of consistency relation in the squeezed limit and the approximation $k'=|\mf{k+q}|\simeq k$ are valid. The contributions from the $\Re\psi_\text{2,bulk}$ and $\Im\psi_\text{2,bulk}$ are the first term and the second term, respectively. We kept the leading-order terms in slow-roll for each contribution. The first term involves the famous factor $(1-n_s)^2$, which is proportional to the square of the squeezed bispectrum of $\zeta$. 
The result \eqref{bulk_inf} shows that the leading contribution from $\psi_\text{2,bulk}$ is proportional to $1/\epsilon$.  
From the counting of the number of $\epsilon$, this term must come from the cubic boundary terms of order $\epsilon$ which were not considered in the evaluation of decoherence rate in previous works (see e.g., \cite{Arroja:2011yj} for an explicit form of such vertices).

Note that we also have the cross term between $\Im\psi_\text{2,bulk}$ and $\Im\psi_\text{2,bd}$, which can be evaluated in a straightforward manner. We do not show them here since they are subdominant.

We reconfirm the dominance of boundary term found in \cite{Sou:2022nsd}; however, there is a quantitative difference. In contrast to our result \eqref{bd_inf}, the result of \cite{Sou:2022nsd} (see also \cite{Martin:2018zbe}) involves an enhancement factor $\Delta N$ which denotes the number of e-folds from the beginning of inflation to the horizon crossing of the system modes $\pm \mf{q}$. This factor comes from the regularization of IR divergence by introducing an IR cutoff at the horizon scale at the beginning of inflation. This could be a quite huge number; in fact, the number $\Delta N\sim 10^4$ is used in \cite{Sou:2022nsd}. If it is very large to compensate for the smallness of the prefactor of \eqref{bd_inf}, the purity is almost lost for any mode that crossed the horizon during inflation. 

\subsection{Evolution after inflation}\label{sec:after}
In this subsection, we briefly discuss decoherence for superhorizon fluctuations after the end of inflation, until just before the horizon re-entry. We consider the instant transition from inflationary epoch to the radiation-dominated (RD) era at the conformal time $\tau_\text{e}$.
For simplicity, we model the evolution of $\zeta$ at superhorizon scales by the EoM of massless scalar field, just like the gravitational waves. In this case, 
the mode function of $\zeta$ during the RD era is (see e.g.,~\cite{Maggiore:1999vm})
\begin{align}
    u_k^\zeta(\tau\geq\tau_\text{e})
    \approx
    \frac{H_\text{e}}{\Mpl\sqrt{2\epsilon}}
    \frac{-i}{a(\tau)}
    \left[
        \alpha_k\frac{e^{-ik\tau}}{\sqrt{2k}} 
        +
        \beta_k\frac{e^{ik\tau}}{\sqrt{2k}}
    \right]
    \qquad
    \text{(RD era)}
\end{align}
with
\begin{align}
    \alpha_k=1-\frac{i}{k\tau_\text{e}}-\frac{1}{2(k\tau_\text{e})^2}
    \,,\quad
    \beta_k=\frac{e^{-2ik\tau_\text{e}}}{2(k\tau_\text{e})^2}
    \,,\quad
    a(\tau\geq\tau_\text{e})
    =
    \frac{\tau-2\tau_\text{e}}{H_\text{e}\tau_\text{e}^2}
    \qquad
    \text{(RD era)}
    \,.\label{connection}
\end{align}
Here, we assumed that the Hubble parameter $H_\text{e}$ and the slow-roll parameter $\epsilon$ are approximately constant during inflation. 
Since this model captures important properties, such as the conservation of $\zeta$ and the $\mathcal{O}((k/aH)^2)$-suppression of its time derivative, we expect that our results capture the essential features of decoherence of $\zeta$ after the end of inflation. 

In this model, the quadratic wavefunction coefficients evolve as
\begin{align}
    \psi_\text{2,bd}(k;\tau)
    &\simeq
    18iH(\tau)\Mpl^2a^3(\tau)
    \qquad
    (k\tau\ll 1)\,,\\
    \Im\psi^\text{RD}_\text{2,bulk}(k;\tau)
    &\simeq
    2 \epsilon\Mpl^2 H_\text{e} (k\tau_\text{e})^2\frac{a^3(\tau)}{3}
    \qquad
    (k\tau\ll1\,,\,\,\tau\gg -\tau_\text{e})
    \,,\\
    \Re\psi^\text{RD}_\text{2,bulk}(k;\tau)
    &\simeq \Re\psi_\text{2,bulk}(k;\tau_\text{e})
    \qquad
    (k\tau\ll 1\,,\,\, \tau\gg -\tau_\text{e})
    \,.
\end{align}
After the end of inflation, $\dot u^\zeta$ stops decaying and remains constant. As a result, $\Im\psi^\text{RD}_\text{2,bulk}$ grows in proportion to $a^3$. By contrast, $\Re\psi^\text{RD}_\text{2,bulk}$ remains constant due to the conservation of $\zeta$. Hence the contribution of $\psi_\text{2,bulk}^\text{RD}$ to $\overline{\Delta}^g_q$ is dominated by 
$\Im\psi_\text{2,bulk}^\text{RD}$ and grows rapidly during the radiation-dominated era:
\begin{align}
    \eval{\overline{\Delta}^g_q(\tau>\tau_\text{e})}_{k<aH,\psi_\text{2,bulk}^\text{RD}}
    &\approx
    2P_\zeta(q;\tau)
    \int_q^{1/\tau}\frac{\mathrm{d}k}{2\pi^2}k^2 
    P_\zeta^2(k;\tau)
    \left|
    \left(3-k\frac{\der}{\der k}\right)
    \Im\psi_\text{2,bulk}^\text{RD}(k;\tau)
    \right|^2
    \no\\
    &\approx
    \frac{H_\text{e}^2}{144 \pi ^2\Mpl^2 \epsilon }
    \frac{1}{(-q\tau_\text{e})^{3}}(a/a_\text{e})^5
    \,.
\end{align}
Here, we assumed $\tau\gg |\tau_\text{e}|$ and used the approximation $a(\tau)H(\tau)\simeq 1/\tau$ which is valid when $\tau\gg |\tau_\text{e}|$.
The integral is dominated by the upper end $k\sim 1/\tau$, and hence the use of consistency relation in the squeezed limit and the approximation $k'=|\mf{k{+}q}|\simeq k$ is valid. 
The power of $(a/a_\text{e})$ is not six but five, as the upper limit of the integral decreases as $(a_\text{e}/a)$, leading to a suppression by a single power.  

By contrast, the growth rate of the boundary term $\Im\psi_\text{2,bd}$ is proportional to the scale factor $a$ during the radiation-dominated era, since $Ha$ decays in this epoch. Hence, the boundary contribution to $\overline{\Delta}_q$ grows by a factor $(a/a_\text{e})^2$ after $\tau>\tau_\text{e}$:
\begin{align}
    \eval{\overline{\Delta}^g_q(\tau>\tau_\text{e})}_{k<aH,\psi_\text{2,bd}}
    \approx
    \eval{\overline{\Delta}^g_q(\tau_\text{e})}_{k<aH,\psi_\text{2,bd}}(a/a_\text{e})^2
    \approx
    \frac{729(16 \ln2-9) H_\text{e}^2}{256 \pi ^2  \Mpl^2\epsilon ^3}\frac{1}{(q\tau_\text{e})^6}(a/a_\text{e})^2
     \,.
\end{align}
We find that the dominant contribution still comes from the $\Im\psi_\text{2,bd}$ when $q$ is at superhorizon scales, as in the case of decoherence during inflation. When $q$ is about to re-enter the horizon and we have $q\sim 1/\tau$, the decoherence rate induced by the boundary contribution is
\begin{align}
    \eval{\overline{\Delta}^g_q(\tau\sim 1/q)}_{k<aH,\psi_\text{2,bd}}
    \sim
    \frac{729(16 \ln2-9) H_\text{e}^2}{256 \pi ^2  \Mpl^2\epsilon ^3}\frac{1}{(q\tau_\text{e})^8}
    \sim
    0.6\times \frac{H_\text{e}^2}{\Mpl^2\epsilon^3}
    \frac{1}{(q\tau_\text{e})^8}
    \,,\label{bd_RD}
\end{align}
showing the scaling $\sim (q\tau_\text{e})^{-8}$. Interestingly, the bulk contribution is still subdominant but exhibits the same scaling
\begin{align}
    \eval{\overline{\Delta}^g_q(\tau\sim 1/q)}_{k<aH,\psi_\text{2,bulk}}
    \sim
    \frac{H_\text{e}^2}{144 \pi ^2\Mpl^2\epsilon }
    \frac{1}{(q\tau_\text{e})^8}
    \sim
    0.0008\times \frac{H_\text{e}^2}{\Mpl^2\epsilon}
    \frac{1}{(q\tau_\text{e})^8}
    \,,\label{bulk_RD}
\end{align}
though a prefactor is smaller than that of the dominant boundary contribution. This scaling implies $\overline{\mathcal{P}}_\sys^g\propto (|\beta_q|^{-2})^2$ where $\beta_q$ is given in \eqref{connection}. Note that this is the same behavior as the one found in \eqref{excited} in a different toy model. This scaling is natural, since the purity for a single system will be proportional to the inverse of occupation number $|\beta_q|^2$ of environment with which the system is entangled, and the additional power $2$ is explained by the number of system modes $\pm \mf{q}$. 

Using \eqref{bd_RD}, we estimate condition for the size of $|q\tau_\text{e}|$ for which the quantum coherence could remain by solving $\eval{\overline{\Delta}^g_q(\tau\sim 1/q)}_{k<aH,\psi_\text{2,bd}}\leq 0.1\ll1$. The solution is
\begin{align}
    |q\tau_\text{e}|
    \geq 
    0.13\times 
    \left(\frac{H_\text{e}}{10^{13}\,\,\text{GeV}}\right)^{1/4}
    \left(\frac{0.1}{\epsilon}\right)^{3/8}
    \,,\label{limit}
\end{align}
meaning that coherence could be preserved only for the modes that crossed the horizon within $2$ e-folds before the end of inflation when $H_\text{e}=10^{13}$ GeV and $\epsilon=0.1$. As explained at the end of Section~\ref{sec:before}, if there were an enhancement factor such as $\Delta N\gg1$, this window may be closed. This highlights the practical importance of considering effective quantum state and its purity.

\section{Conclusion}\label{conclusion}
We studied the quantum decoherence of the curvature perturbations $\zeta$ at a certain superhorizon scale $\pm \mf{q}$ in the canonical single field inflation model. 
In this model, the decoherence is caused by the nonlinear gravitational interactions between different scales. 
In particular, we tackled the issues of IR and UV divergences that were reported in the literature to exist in the decoherence rate computed using the reduced density matrix. If these divergences were really relevant to the system, the coherence of the system would be significantly broken by the environment at very different scales.

We calculated the purity for the superhorizon modes $\pm\mf{q}$ at one-loop level by formally taking all the cubic interactions into account through the consistency relation which follows from the spatial diffeomorphism invariance. This treatment considerably simplifies the one-loop computations. We confirmed that the purity suffers from IR and UV divergences. Crucially, these divergences appear in the non-separable piece of the reduced density matrix, which are not simply removable by the usual renormalization. 

The main observation of this paper is that these divergences will be absent in actual observables measured by a local observer, and hence there should exist a proper measure of the coherence of the system which is free from divergences. 
Therefore, we proposed to consider an \textit{effective quantum state}, defined in terms of actual observables, as a more appropriate probe of the quantum coherence of the system measured by a local observer. As explained in Section~\ref{sec:eff_state}, our effective reduced density matrix (or equivalently, an effective Wigner function) is designed so that it reproduces all the correlators measured by a local observer. Then, we proposed to use an effective purity defined by our effective state as a probe of genuine coherence of the system. An advantage of our approach is that it is straightforward to incorporate actual observables into the study of quantum state.  

Subsequently, we applied our effective state formalism and obtained a manifestly IR and UV finite effective purity. {We emphasize that no artificial cutoffs are needed in our approach. Once the effective state is properly defined, the resulting effective purity is manifestly finite; this is achieved solely through standard procedures such as conventional renormalization, without altering any prescriptions.}

For the IR side, we adopted the correlators which are characterized by the geodesic dinstance proposed in~\cite{Urakawa:2009my,Urakawa:2010kr,Tanaka:2011aj,Tanaka:2012wi,Tanaka:2013xe,Pajer:2013ana,Tanaka:2013caa,Tanaka:2014ina,Tanaka:2017nff} as proper observables measured by a local observer. We also highlighted that the consistency relation plays an important role for the cancellation of IR divergences in the computation of effective purity. Concretely, we demonstrated the cancellation of IR contributions in the cubic wavefunction coefficient in geodesic coordinate. Because of the cancellation, the deep IR modes do not spoil the coherence of the system.
This result is suggestive since it may indicate the factorization of the quantum state $\rho^g=\rho^g_{\text{IR}}\rho^g_{\text{non-IR}}$. It would be valuable to clarify the entanglement structure of deep IR sector by properly treating Jacobian from the field transformation~\eqref{eq:jaco}. 

We also addressed a concern for the decoherence from the bulk term 
raised in~\cite{Nelson:2016kjm}: the decoherence of superhorizon modes caused dominantly by the environment at shorter scales may be merely a gauge artifact and may be canceled in local observables. We clarified that the cancellation does not occur, in contrast to the case where decoherence of the system modes is caused by a superhorizon environment at scales much larger than that of the system.

For the UV side, it was suggested in~\cite{Agon:2014uxa,Agon:2017oia,Bucciotti:2024lvb} that the UV divergences originate from the equal-time limit of correlators in momentum space, and the finite time resolution will be important for ensuring the finiteness of such correlators~\cite{Bucciotti:2024lvb,Burgess:2024heo}. We considered the correlators of time-averaged operators as observables, consistently with these works.
We demonstrated that the time smearing provides a robust mechanism for eliminating the spurious UV-divergent quantum decoherence of the system caused by the violent zero-point fluctuations of the deep UV modes. This is consistent with the statement of ``false decoherence'' proposed in \cite{unruh:2011fal}. 
Intuitively, the cancellation occurs since the time smearing effectively projects out the high frequency components contained in operators $\zeta_{\pm \mf{q}}(\tau)$ and $\pi_{\pm\mf{q}}(\tau)$ which are local in time. 
On the other hand, we found that an excited environment, such as the modes which crossed the horizon during inflation, serves as a source of genuine decoherence, due to the phase cancellations of oscillations. {We demonstrated this point using a Bogoliubov coefficient in a toy model in Section~\ref{sec:eg}.} An important lesson we learned is that large entanglement predicted by the reduced density matrix does not necessarily imply a loss of coherence of the system.

Our result also applies to purity lost due to heavy physics in flat space; the violation of unitary time evolution of the system due to the ignorance of heavy physics beyond the EFT resolution (in both space and time) will be exponentially suppressed, as long as the excitations of heavy particles are negligible. This is consistent with the observation made in~\cite{Burgess:2024heo}. Note that this issue is absent in the S-matrix, since the on-shell condition automatically suppresses the high-frequency modes for a given three-momentum. 

In Section~\ref{sec:finite_deco}, we evaluated the effective purity {quantitatively.}
In particular, we focused on decoherence caused by the superhorizon environment during inflation or during the subsequent radiation-dominated era. We assumed the instantaneous reheating scenario, and we modeled the time evolution of curvature perturbations at superhorizon scales after inflation by the EoM of massless scalar field to simplify the analysis. We solved the time evolution until the system mode $q$ about to re-enter the horizon. Our calculations account for all the relevant cubic interactions since the consistency relation \eqref{eq:psi2} is used appropriately. 

It was reported in \cite{Sou:2022nsd} that the cubic boundary vertex $\Lag\ni\del_t(9\mpl^2Ha^3\zeta^3)$ is the dominant source of decoherence during inflation. We reconfirmed this, and moreover, we found that it remains dominant up to when the system re-enters the horizon. The contributions from the cubic vertices predicted by the quadratic bulk term via \eqref{eq:psi2} are always sub-dominant; during inflation, it is much smaller than the boundary contribution \eqref{bd_inf}. However, near the horizon re-entry, the bulk contribution \eqref{bulk_RD} and the boundary contribution \eqref{bd_RD} exhibit the same scaling behavior $\sim (q\tau_\text{e})^{-8}$, which has a clear interpretation as explained below \eqref{bulk_RD}. Moreover, in contrast to the result obtained in \cite{Sou:2022nsd}, our result is not enhanced by the duration of inflation because our effective purity is not influenced by the deep IR modes. Since the duration of inflation could be very long, this highlights a practical importance of considering our effective purity. 

Finally, we derived an upper bound on the scales for which curvature perturbations maintain coherence at horizon re‑entry (see \eqref{limit}). For typical parameter values, coherence is preserved only by modes that crossed the horizon within two e‑folds before the end of inflation.

It would be straightforward and interesting to extend our analysis to the case of gravitational waves decohered by gravitational non-linear interactions. As in the case of $\zeta$, the modes that crossed the horizon within a few e-folds before the end of inflation would retain coherence. It would be valuable to follow the time evolution after the horizon re-entry in detail, and precisely evaluate the remaining coherence in actual observables. It would be also important to study decoherence induced by the thermal environment (as recently addressed in \cite{Takeda:2025cye}). We leave these aspects for future work.

\section*{Acknowledgment}
We would like to thank Robert Brandenberger, Simon Caron-Huot, Amaury Micheli, Shintaro Minagawa, Chon Man Sou, Yuichiro Tada, Takahiro Tanaka, Junqi Wang, and Yi Wang for their helpful discussions. We also thank Takeshi Kobayashi, Dimitrios Kranas, Francescopaolo Lopez, J\'er\^ome Martin, and Vincent Vennin for their fruitful discussions and kind hospitality when F.S. visited their institute during the project.
F.S. acknowledges financial aid from the Institute for Basic Science under the project code IBS-R018-D3, and JSPS Grant-in-Aid for Scientific Research No. 23KJ0938. J.T. was supported by the JSPS Postdoctoral Fellowships for Research Abroad.


\addcontentsline{toc}{section}{References}
\bibliographystyle{setting/JHEP}
\bibliography{refs}

\newpage
\appendix

\section{One-loop corrections}\label{sec:one-loop}
\subsection{Cancellation of growing terms}\label{sec:nogrow}

Here we estimate $\Re A -C$ and it does not grow at a late time.
The presence of growing terms in $\Re A$ and $C$ is 
a generic feature found in various theories with light fields in inflationary spacetime. 
Therefore, we first evaluate $\Re A -C$ in $g\phi^3$ theory on dS background and show the cancellation of the dominant growing terms. We then extend the argument to the case of curvature perturbation. 

We assume that the mass of $\phi$ is much lighter than the Hubble scale so that the mode function can be approximated by the massless one for simplicity. However, this approximation does not change the final conclusion. $\Re A$ and $C$ in this model are evaluated at one-loop level as 
\begin{subequations}
    \label{cancel_eg}
\begin{align}
    &\eval{\Re A}_\text{one-loop, $g\phi^3$}
    =
    -|u_q(\tau)|^{-4}
    \left[ \int_{\tau',\tau''} u_q(\tau)u_q^*(\tau')u_q^*(\tau'')u_q(\tau)iV_{++}(\tau',\tau'';\mf{q})
    + \text{(c.c)}
    \right]
    \,,\\
    &\eval{C}_\text{one-loop, $g\phi^3$}
    =
    |u_q(\tau)|^{-4}
    \left[ \int_{\tau',\tau''} u_q(\tau)u_q^*(\tau')u_q(\tau'')u_q^*(\tau)iV_{+-}(\tau',\tau'';\mf{q})
    + \text{(c.c)}
    \right]
    \,,
\end{align}
\end{subequations}
where $\int_{\tau',\tau''}\equiv \int^{\tau}_{-\infty}\mathrm{d}\tau'\,a^4(\tau')\int^{\tau}_{-\infty}\mathrm{d}\tau''a^4(\tau'')$ and
$V_{\sigma\sigma'}$ with $\sigma,\sigma'=\pm$ are given at one-loop level by
\begin{align}
    iV_{\sigma\sigma'}(\tau',\tau'';\mf{q})
    =
    \frac{1}{2}
    \text{sgn}[\sigma\sigma'] \,g^2
    \int_{\mf{k}, k> q}
    G_{\sigma\sigma'}(\tau',\tau'';k)G_{\sigma\sigma'}(\tau',\tau'';|\mf{k+q}|)
    \,.\label{Vsigma}
\end{align}
Here, we ignored an irrelevant contribution from a diagram which is identical to the rightmost diagram in Fig.~\ref{fig:quartic}. $u_q$ denotes the mode function for a free scalar field, and $G_{\sigma\sigma'}(\tau,\tau';k)$ denotes the conventional free-theory two-point correlation functions of $\sigma$ and $\sigma'$-fields in the Schwinger-Keldysh formalism; for more details, see \cite{Kamenev:2009jj}.

It is easy to extract the dominant pieces at a late time by rewriting \eqref{cancel_eg} into the following form, which is naturally motivated by the expression in the Keldysh basis,
\begin{subequations}
    \label{cancel_eg2}
\begin{align}
    -|u_q(\tau)|^{4}\eval{\Re A}_\text{one-loop, $g\phi^3$}
    &=
    \int_{\tau',\tau''}
    \biggl\{
    \re[u_q(\tau)u_q^*(\tau')]\re[u_q^*(\tau'')u_q(\tau)]
    \left[iV_{++}+iV_{--}\right]
    \no\\
    &\hspace{8mm}
    +
    i\re[u_q(\tau)u_q^*(\tau')]\im[u_q^*(\tau'')u_q(\tau)]
    \left[iV_{++}-iV_{--}\right]
    \no\\
    &\hspace{8mm}
    +
    i\im[u_q(\tau)u_q^*(\tau')]\re[u_q^*(\tau'')u_q(\tau)]
    \left[iV_{++}-iV_{--}\right]
    \no\\
    &\hspace{8mm}
    -
    \im[u_q(\tau)u_q^*(\tau')]\im[u_q^*(\tau'')u_q(\tau)]
    \left[iV_{++}+iV_{--}\right]
    \biggr\}
    \,,\label{cancel_eg2_A}\\
    |u_q(\tau)|^{4}\eval{C}_\text{one-loop, $g\phi^3$}
    &=
    \int_{\tau',\tau''}
    \biggl\{
    \re[u_q(\tau)u_q^*(\tau')]\re[u_q^*(\tau'')u_q(\tau)]
    \left[iV_{+-}+iV_{-+}\right]
    \no\\
    &\hspace{8mm}
    +
    i\re[u_q(\tau)u_q^*(\tau')]\im[u_q^*(\tau'')u_q(\tau)]
    \left[-iV_{+-}+iV_{-+}\right]
    \no\\
    &\hspace{8mm}
    +
    i\im[u_q(\tau)u_q^*(\tau')]\re[u_q^*(\tau'')u_q(\tau)]
    \left[iV_{+-}-iV_{-+}\right]
    \no\\
    &\hspace{8mm}
    +
    \im[u_q(\tau)u_q^*(\tau')]\im[u_q^*(\tau'')u_q(\tau)]
    \left[iV_{+-}+iV_{-+}\right]
    \biggr\}
    \,,\label{cancel_eg2_C}
\end{align}
\end{subequations}
where  we suppressed the arguments of $V_{\sigma\sigma'}$ for simplicity.

The growing pieces at a late time come from the integral around $\tau',\tau''\sim \tau$ in \eqref{cancel_eg2}, where the mode $\mf{q}$ is always at superhorizon. In this regime, the mode function is effectively frozen and dominated by the constant piece, leading to 
\begin{align}
    \re[u_q(\tau')u_q^*(\tau'')] \sim q^{-3}
    \,,\quad
    \im[u_q(\tau')u_q^*(\tau'')] \sim \tau'^{3}-\tau''^{3}
    \,,\label{mode_super}
\end{align}
where we ignore irrelevant prefactors. Hence, the dominant terms in $\Re A$ and $C$ are the terms in the first line of \eqref{cancel_eg2_A} and \eqref{cancel_eg2_C}, respectively. 

The dominant term in $\Re A- C$ is written as the integral of $V_{cc}(\tau',\tau'')$, where $V_{cc}\equiv V_{++}+V_{--}+V_{+-}+V_{-+}$. However, $V_{cc}(\tau',\tau'' )$ vanishes since the vertices $\tau'$ and $\tau'' $ are connected by both of the retarded Green's function $G_{c\Delta}$ and the advanced Green's function $G_{\Delta c}$ and it is impossible to satisfy the condition $\tau'>\tau''$ and $\tau'<\tau''$ simultaneously; 
\begin{align}
    V_{cc}(\tau',\tau'')
    \propto
    g^2\int_\mf{k} G_{c\Delta}(\tau',\tau'';k)G_{\Delta c}(\tau',\tau'';|\mf{k}+\mf{q}|)
    =0\,.\label{cc}
\end{align}

Let us estimate the remaining sub-leading terms. {They are given by the sum of diagrams with at least two retarded Green's functions. For the superhorizon environment $q<k<|\tau|^{-1}$, as shown in \eqref{mode_super}, the retarded Green's function decays rapidly while the Wightman function is constant in time. This verifies that the sub-leading terms in $\Re A$ and $C$ do not grow in time as fast as inverse powers of $\tau$, when the superhorizon environment $q<k<|\tau|^{-1}$ is considered.} 
For the subhorizon environment $|k\tau|\gtrsim 1$, $V_{\sigma\sigma'}$ decays rapidly for $H|\tau'-\tau''|\gg 1$ due to the rapid oscillation, and it exhibits the following simple scaling for $\tau'\sim\tau''$ since the time dependence of the mode function with the fixed physical momentum $k_\text{ph}\equiv -Hk\tau'$ is $u_k(\tau')=\tau'^{3/2}\times \text{[function of $k_\text{ph}]$}$:
\begin{align}
    \eval{V_{\sigma\sigma'}(\tau',\tau'')}_{\tau'\sim \tau'', -k\tau'\gtrsim H}
    =
    \frac{1}{2}\text{sgn}[\sigma\sigma'] \,g^2
    \int_{\mf{k}, -k\tau'\gtrsim 1\gg -q\tau'}
    G_{\sigma\sigma'}(\tau',\tau'';k)G_{\sigma\sigma'}(\tau',\tau'';|\mf{k+q}|)
    \sim \tau'^{3}
    \,.
\end{align}
Combining this with Eq.~\eqref{mode_super}, we find that the sub-leading terms in \eqref{cancel_eg2} cannot grow as fast as inverse powers of $\tau$. 
Thus, we conclude that $\Re A -C$ does not contain the terms which grow at a late time due to the cancellation \eqref{cc}.

In the case of $\zeta$, the form of cubic couplings are more involved and it is hard to evaluate $V_{\sigma\sigma'}$ explicitly. Moreover, they involve derivative interactions. A contribution from each diagram with time derivatives is schematically given by the RHS of \eqref{cancel_eg2} with time derivatives acting on the integrand. This gives the additional factors which are some non-negative powers of the physical momentum $q_\text{ph}$ or $k_\text{ph}$ at the time $\tau'\sim\tau''$, compared to the case without derivatives. The same holds in the case with spatial derivatives, since they are always accompanied by the appropriate number of scale factors. {Therefore, as expected, the late-time behavior of $\Re A -C$ cannot be more singular than in the case without derivative interactions.} 
We conclude that $\Re A -C$ cannot grow as fast as inverse powers of $\tau$ at a late time. 
Note that the cubic couplings also depend on time, strictly speaking, but it will not matter as long as the size of couplings remain tiny. 

\subsection{One-loop corrections from bubble diagrams}\label{sec:LoopRecord}
Several amplitudes of one-loop diagrams shown in Fig.~\ref{fig:oneloop} are recorded. Explicitly, the contributions from the integration range $\tau'<\tau_2$ are given by
\begin{align}
    &F_{+-}(\tau_1,\tau_2)
    =
    \int^{\tau_1}_{-\infty}\mathrm{d}\tau
    \int^{\tau_2}_{-\infty}\mathrm{d}\tau'\,a^4(\tau)a^4(\tau')
    \sum_{A,B=\zeta,\pi}
    {u_q^A}^*(\tau) u_q^B(\tau') 
    iV^{AB}_{+-}(\tau,\tau')
    \,,\\
    &F_{-+}(\tau_1,\tau_2)
    =
    \int^{\tau_1}_{-\infty}\mathrm{d}\tau
    \int^{\tau_2}_{-\infty}\mathrm{d}\tau'\,a^4(\tau)a^4(\tau')
    \sum_{A,B=\zeta,\pi}
    {u_q^A}(\tau) {u_q^B}^*(\tau') 
    iV^{AB}_{-+}(\tau,\tau')
    \,,\\
    &F_{++}(\tau_1,\tau_2)
    =
    \int^{\tau_1}_{-\infty}\mathrm{d}\tau
    \int^{\tau_2}_{-\infty}\mathrm{d}\tau'\,a^4(\tau)a^4(\tau')
    \sum_{A,B=\zeta,\pi}
    {u_q^A}^*(\tau) {u_q^B}^*(\tau') 
    iV^{AB}_{++}(\tau,\tau')
    \,,\\
    &F_{--}(\tau_1,\tau_2)
    =
    \int^{\tau_1}_{-\infty}\mathrm{d}\tau
    \int^{\tau_2}_{-\infty}\mathrm{d}\tau'\,a^4(\tau)a^4(\tau')
    \sum_{A,B=\zeta,\pi}
    {u_q^A}(\tau) u_q^B(\tau') 
    iV^{AB}_{--}(\tau,\tau')
    \,,
\end{align}
Here, $V^{AB}_{\sigma\sigma'}$ denotes the amplitude of the diagrams shown in Fig.~\ref{fig:oneloop} after amputating their external lines. They are computed as the loop integral of the product of two internal propagators (with possible derivatives acting on them), multiplied by the appropriate coupling constants. The indices $A$ and $B$ label the types of field ($\zeta$ or $\pi$) on the external legs at the vertices $\tau$ and $\tau'$, respectively. As mentioned in the main text, $F_{+-}$ and its complex conjugate $F_{-+}$ can be expressed using the tree-level cubic wavefunction coefficient \eqref{F_wavefn}.

The contributions of $\tau'>\tau_2$ are 
\begin{align}
    &\Delta I^{\alpha\beta}_{+-}
    =
    u_q^\alpha(\tau_1){u_q^\beta}^*(\tau_2)
    \Delta F_{+-}(\tau_1,\tau_2)
    \,,\\
    &\Delta I^{\alpha\beta}_{-+}
    =
    {u_q^\alpha}^*(\tau_1){u_q^\beta}^*(\tau_2)
    \Delta F_{-+}(\tau_1,\tau_2)
    \,,\\    
    &\Delta I^{\alpha\beta}_{++}
    =
    {u_q^\alpha}(\tau_1){u_q^\beta}^*(\tau_2)
    \Delta F_{++}(\tau_1,\tau_2)
    \,,\\
    &\Delta I^{\alpha\beta}_{--}
    =
    {u_q^\alpha}^*(\tau_1){u_q^\beta}^*(\tau_2)
    \Delta F_{--}(\tau_1,\tau_2)
    \,,
\end{align}
with
\begin{align}
    &\Delta F_{+-}(\tau_1,\tau_2)
    =
    \int^{\tau_1}_{-\infty}\mathrm{d}\tau
    \int^{\tau_1}_{\tau_2}\mathrm{d}\tau'\,a^4(\tau)a^4(\tau')
    \sum_{A,B=\zeta,\pi}
    {u_q^A}^*(\tau) u_q^B(\tau') 
    iV^{AB}_{+-}(\tau,\tau')
    \,,\\
    &\Delta F_{-+}(\tau_1,\tau_2)
    =
    \int^{\tau_1}_{-\infty}\mathrm{d}\tau
    \int^{\tau_1}_{\tau_2}\mathrm{d}\tau'\,a^4(\tau)a^4(\tau')
    \sum_{A,B=\zeta,\pi}
    {u_q^A}(\tau) {u_q^B}(\tau') 
    iV^{AB}_{-+}(\tau,\tau')
    \,,\\
    &\Delta F_{++}(\tau_1,\tau_2)
    =
    \int^{\tau_1}_{-\infty}\mathrm{d}\tau
    \int^{\tau_1}_{\tau_2}\mathrm{d}\tau'\,a^4(\tau)a^4(\tau')
    \sum_{A,B=\zeta,\pi}
    {u_q^A}^*(\tau) {u_q^B}(\tau') 
    iV^{AB}_{++}(\tau,\tau')
    \,,\\
    &\Delta F_{--}(\tau_1,\tau_2)
    =
    \int^{\tau_1}_{-\infty}\mathrm{d}\tau
    \int^{\tau_1}_{\tau_2}\mathrm{d}\tau'\,a^4(\tau)a^4(\tau')
    \sum_{A,B=\zeta,\pi}
    {u_q^A}(\tau) u_q^B(\tau') 
    iV^{AB}_{--}(\tau,\tau')
    \,.
\end{align}
Note that the integrand is finite for finite $\epsilon$ under the usual $i\epsilon$-prescription, which ensures that $\Delta F$ vanishes when the equal time $\tau_2=\tau_1$ is considered.

\section{Derivation of the second order wavefunction coefficient}\label{sec:psi2}
Here, we derive the second order wavefunction coefficient $\psi_2$ in the expansion of the wavefunction at the boundary $\zeta=\bar{\zeta}$ and $\tau=\bar{\tau}$,
\begin{align}
    \Psi[\bar{\zeta};\bar{\tau}]\equiv \braket{\bar{\zeta}|\psi(\bar{\tau})}
    =\exp[-\frac{1}{2}\int_{\mf{k}}\bar{\zeta}_{\mf{k}}\bar{\zeta}_{-\mf{k}}\psi_2(k)-\frac{1}{6}\int_{\mf{k},\mf{k}'}\bar{\zeta}_{\mf{k}}\bar{\zeta}_{\mf{k}'}\bar{\zeta}_{-\mf{k}-\mf{k}'}\psi_3(k,k',|\mf{k}+\mf{k}'|)+\cdots],
\end{align}
by using the semiclassical approximation
\begin{align}
    \Psi[\bar{\zeta};\bar{\tau}]=\int_{\mr{BD}}^{\bar{\zeta}}\mathscr{D}\zeta\ e^{iS}\approx e^{iS_\mr{cl}[\bar{\zeta}]}.
\end{align}
For the purpose of calculating bispectrum using Maldacena's consistency relation, it is helpful to use Mukhanov--Sasaki variables
\begin{align}
    u_\mf{k}\equiv -z\zeta_\mf{k},\qquad z\equiv \frac{a\dot{\phi}}{H},
\end{align}
which leads to the following second order action for curvature perturbations
\begin{align}
    S_\mr{cl}^{(2)}=\int d\tau\frac{d^3\mf{k}}{(2\pi)^3}\frac{1}{2}\qty[|u'_\mf{k}|^2-k^2|u_\mf{k}|^2+\frac{z''}{z}|u_\mf{k}|^2-\frac{\del}{\del\tau}\qty(aH\qty(1+\frac{\eta}{2}) {}|u_\mf{k}|^2+\frac{9aH}{\epsilon}|u_\mf{k}|^2-\frac{k^2}{aH\epsilon}|u_\mf{k}|^2)],
\end{align}
where we kept all of time boundary terms. Note that the first boundary term arises because of using Mukhanov--Sasaki variable. 
By solving the equation of motion 
\begin{align}
    u_\mf{k}''+\qty(k^2-\frac{z''}{z})u_\mf{k}=0
\end{align}
with including the first order of slow-roll parameters
\begin{align}
    \frac{z''}{z}\simeq \frac{1}{\tau^2}\qty(2+3\qty(\nu-\frac{3}{2})),\qquad \nu\equiv\frac{3}{2}-\frac{2\epsilon+\eta}{2},
\end{align}
and with the boundary conditions $u_\mf{k}(-\infty)=0$, $u_\mf{k}(\tau)=\bar{u}_\mf{k}=-z\bar{\zeta}_\mf{k}$, we derive the bulk-to-boundary propagator
\begin{align}
    u_\mf{k}(\tau)\equiv K_\mf{k}(\tau)\bar{u}_\mf{k},\qquad K_\mf{k}(\tau)=\sqrt{\frac{\tau}{\bar{\tau}}}\frac{H_\nu^{(2)}(-k\tau)}{H_\nu^{(2)}(-k\bar{\tau})},
\end{align}
where $H_\nu^{(2)}(z)$ is the Hankel function of the second kind.\footnote{{The Bunch--Davies initial condition is used in $u_\mf{k}(-\infty)=0$ with $i\epsilon$ prescription. For general $\alpha$-vacua, both positive and negative frequency modes exist and we cannot impose the boundary condition.}}
By substituting the solution to the classical action and using the equation of motion,
\begin{align}
    S_\mr{cl}^{(2)}=\int\frac{d^3\mf{k}}{(2\pi)^3}a^2k\mpl^2|\bar{\zeta}_\mf{k}|^2\qty[-\epsilon\frac{H_{\nu-1}^{(2)}(-k\bar{\tau})}{H_\nu^{(2)}(-k\bar{\tau})}-\frac{9aH}{k}+\frac{k}{aH}]
\end{align}
is obtained, and thus $\psi_2(k)$ reads
\begin{align}
    \psi_2=\frac{2i\mpl^2k^3}{H^2} \qty(\epsilon\qty(\frac{aH}{k})^2\frac{H_{\nu-1}^{(2)}(-k\bar{\tau})}{H_\nu^{(2)}(-k\bar{\tau})}+9\qty(\frac{aH}{k})^3-\frac{aH}{k}).
\end{align}

\section{Reviews of IR behavior in local observables}\label{sec:loe}
\subsection{Curvature perturbation}
In this Appendix, we briefly review IR behavior for local observers based on earlier works~\cite{Urakawa:2009my,Urakawa:2010kr,Tanaka:2011aj,Tanaka:2012wi,Tanaka:2013xe,Pajer:2013ana,Tanaka:2013caa,Tanaka:2014ina,Tanaka:2017nff}. 
The comoving coordinate is not geodesic distance, so the correlation functions such as $\braket{\zeta(\mf{x}_1)\zeta(\mf{x}_2)\zeta(\mf{x}_3)}$ is not a geometrical object as well. What we observe, however, can be written in terms of geometrical object.
To address the subtlety, we introduce the geodesic coordinate $\mf{x}^g$ which expresses the conformally free-falling observer~\eqref{eq:free_fall}.
Under the coordinate transformation, the curvature perturbations measured in the geodesic coordinate are written as
\begin{align}
    \zeta(\mf{x})\longrightarrow \zeta^g(\mf{x}^g)=
    \zeta(\mf{x}(\mf{x}^g))
    \simeq
    \zeta(\mf{x}^g)+\delta{x}^i\pdif{x^{i,g}}\zeta(\mf{x}^g)+\order{\delta{{x}}^2},
\end{align}
where $\delta x^i=x^i-x^{i,g}$.
The formal procedure for introducing the coordinate transformation can be found in the literature shown above. Instead of following the full derivation, it is enough to perform the coordinate transformation $\mf{x}^g= e^{\zeta_\text{L}(\mf{x})}\mf{x}$ to see the modification to IR behavior. 
Here, $\zeta_\text{L}(\mf{x})$ is formally defined in~\eqref{eq:zeta_L}.
From the coordinate transformation, $\delta{x}^i\simeq -\zeta_\text{L}(0){x}^{i,g}$ is satisfied around the origin where the observer is located. Here, the constant mode $\zeta_\text{L}(0)$ includes only long modes and works as $\zeta_\text{L}$.
After Fourier transformation, we obtain~\eqref{eq:loe_corr}.
As a consequence, the interactions between long modes and short modes are modified.
Let us explicitly check the modification.
It should be noted that substituting Eq.~\eqref{eq:loe_corr} to $\braket{\zeta_1^g\zeta_2^g\zeta_3^g}$ easily leads to wrong evaluation due to delta functions. Rather, starting from coordinate space instead of Fourier space makes calculation clearer.
By expanding $\zeta^g(\mf{x}^g)$ in terms of $\zeta(\mf{x}^g)$, the correlation function is written as
\begin{align}
    \braket{\zeta^g(\mf{x}^g_1)\zeta^g(\mf{x}^g_2)\zeta^g(\mf{x}^g_3)} =&\braket{\zeta(\mf{x}^g_1)\zeta(\mf{x}^g_2)\zeta(\mf{x}^g_3)}\notag\\
    &-\qty(x_1^{i,g}\pdif{x_1^{i,g}}+x_2^{i,g}\pdif{x_2^{i,g}}+x_3^{i,g}\pdif{x_3^{i,g}}) \braket{\zeta(\mf{x}^g_1)\zeta(\mf{x}^g_2)\zeta(\mf{x}^g_3)\zeta_\text{L}(0)}
    \,.
\end{align}
After Fourier transformation, the bispectrum in squeezed limit reads
\begin{align}
    \lim_{k_3\to 0}\braket{\zeta_{1}^g\zeta_2^g\zeta_3^g}'=\lim_{k_3\to 0}\qty(\braket{\zeta_{1}\zeta_2\zeta_3}'+\qty(3+k_1\pdif{k_1})\braket{\zeta_1\zeta_{-1}}'\braket{\zeta_3\zeta_{-3}}'),
\end{align}
where we omit the delta function of momentum conservation in $\braket{\cdots}'$.
We obtain the same expression as the one substituting Eq.~\eqref{eq:loe_corr} to only $\zeta^g_{\mf{k}_1}$ of $\braket{\zeta^g_{\mf{k}_1}\zeta^g_{\mf{k}_2}\zeta^g_{\mf{k}_3}}$. It is understood as a consequence of translational symmetry which relates the scale $\mf{k}_1$ to difference of the coordinate $\mf{x}_1-\mf{x}_2$ corresponding to short mode correlations $\braket{\zeta^g_1\zeta^g_2}$, and squeezed bispectrum is the correlation between the short mode two point function and the constant long mode.
In minimal single field inflation, 
\begin{align}
    \lim_{k_3\to 0}\braket{\zeta_{1}^g\zeta_2^g\zeta_3^g}'= \lim_{k_3\to 0}\qty(\braket{\zeta_{1}\zeta_2\zeta_3}'-(1-n_s)\braket{\zeta_1\zeta_{-1}}'\braket{\zeta_3\zeta_{-3}}')
\end{align}
is satisfied, and it exactly cancels with Maldacena's consistency relation.
Furthermore, it is known that next-to-leading order cancels as well, and the genuine leading terms are next-to-next-to-leading order, which correspond to contributions from geometric quantities, i.e., the curvature tensor and the extrinsic curvature.
This cancellation means that the interactions between long modes and short modes are killed in the coordinate. 
Let us consider the correlation function in position space to quickly see how IR divergence is regularized. Fourier transformation gives
\begin{align}
    \braket{\zeta(\mf{x}_1)\zeta(\mf{x}_2)\zeta(\mf{x}_3)}&= \int\frac{d^3\mf{k}_1}{(2\pi)^3}\frac{d^3\mf{k}_2}{(2\pi)^3}\frac{d^3\mf{k}_3}{(2\pi)^3}\braket{\zeta_{1}\zeta_2\zeta_3}e^{-i(\mf{k}_1\cdot\mf{x}_1+\mf{k}_2\cdot\mf{x}_2+\mf{k}_3\cdot\mf{x}_3)}\notag\\
    &\supset\int_{k_1\ll k_2\ll aH}\frac{k_1^2d{k}_1}{(2\pi)^3}\frac{k_2^2d{k}_2}{(2\pi)^3}\frac{1-n_s}{k_1^3k_2^3}\sim (1-n_s)\eval{\log\frac{k_1}{k_2}}_{k_1\to 0}
\end{align}
in comoving coordinate, where we especially picked the contribution from a squeezed limit $k_1\ll k_2,k_3$ in the second line, i.e., Maldacena's consistency relation, and which is logarithmically divergent. In local observables, the power of the momentum changes to the next-next-leading order, so
\begin{align}
    \braket{\zeta^g(\mf{x}_1^g)\zeta^g(\mf{x}_2^g)\zeta^g(\mf{x}_3^g)}_\text{IR}\propto  \int_{ k_1\ll k_2\ll aH}\frac{k_1^2d{k}_1}{(2\pi)^3}\frac{k_2^2d{k}_2}{(2\pi)^3}\frac{1}{k_1k_2^5}
\end{align}
is IR contribution in the geodesic coordinate, and is convergent. 
It should be noted that a similar logarithmic divergence appears in the tree-level power spectrum and is not regularized in local observables. However, since the observable scales are limited to a specific range, the ``IR divergence'' associated with external momenta does not pose a practical problem. Rather, the cancellation in the tree-level bispectrum supports the validity of expanding around a Gaussian distribution in the IR regime.  
The IR finiteness of loop corrections to correlation functions in the Bunch--Davies vacuum has been discussed in the literature~\cite{Urakawa:2009my,Urakawa:2010kr,Tanaka:2012wi,Tanaka:2013xe,Tanaka:2014ina,Tanaka:2017nff}.

\subsection{Conjugate momenta}
Correlation functions involving the conjugate momenta are written as follows:
For tree level two-point correlations,
\begin{align}
    \braket{\{\pi_{\mf{k}_1},\zeta_{\mf{k}_2}\}}'=& \frac{-\Im\psi_{2}(k_1)}{2\Re\psi_{2}(k_1)},\\
    \braket{\pi_{\mf{k}_1}\pi_{\mf{k}_2}}'=& \frac{|\psi_{2}(k_1)|^2}{2\Re\psi_{2}(k_1)},
\end{align}
where $\{\bullet\}$ is Weyl ordering, and for tree level three-point correlations,
\begin{align}
    \label{eq:bis_zzp}
    \braket{\{\pi_{\mf{k}_1},\zeta_{\mf{k}_2},\zeta_{\mf{k}_3}\}}'=& \frac{2\Im[\psi_{2}(k_1)\psi_3^*]}{\prod_{i=1}^32\Re\psi_{2}(k_i)},\\
    \braket{\{\pi_{\mf{k}_1},\pi_{\mf{k}_2},\zeta_{\mf{k}_3}\}}'=& \frac{2\Re[\psi_{2}(k_1)\psi_{2}(k_2)\psi_3^*]}{\prod_{i=1}^32\Re\psi_{2}(k_i)},\\
    \braket{\pi_{\mf{k}_1}\pi_{\mf{k}_2}\pi_{\mf{k}_3}}'=& -\frac{2\Im[\psi_{2}(k_1)\psi_{2}(k_2)\psi_{2}(k_3)\psi_3^*]}{\prod_{i=1}^32\Re\psi_{2}(k_i)}.\label{eq:bis_ppp}
\end{align}
Note that the momentum bispectra also have $1/k_\text{L}^3$ contributions. This is because the quadratic boundary terms give the momenta the contributions proportional to $\zeta$ rather than $\dot{\zeta}$.

The coordinate transformation for conjugate momenta is also discussed in~\cite{Tanaka:2012wi}. 
Considering diffeomorphism invariance of the action, the conjugate momenta should be transformed as $\pi^g(\mf{x}^g)=e^{-3\zeta_\text{L}}\pi(\mf{x})$ by the coordinate transformation. This is actually a canonical transformation which conserves the canonical commutation relation~\eqref{eq:commutation}.
In Fourier space, the momenta transform as Eq.~\eqref{eq:momentum_loe}.
It is formally invariant under fluctuations of deep IR modes by following the formulation in~\cite{Tanaka:2017nff}.
To see the consequence concretely, let us show momentum bispectra in geodesic coordinate. As expected, the comoving bispectra in the squeezed configuration exhibit the $\order{1/k_\text{L}^3}$ behavior, but they cancels out in local observables. When we take $\zeta_k$ to be a long mode, the contributions from the consistency relation exactly cancels in geodesic coordinate, i.e., $\braket{\{\zeta_1^g,\zeta_2^g,\pi_3^g\}}',\braket{\{\zeta_1^g,\pi_2^g,\pi_3^g\}}'\underset{k_1\to 0}{\longrightarrow}0$. When $\pi_k$ is long, the consistency relation is partially canceled and leaves correction terms of $\order{k_\text{L}^0}$,
\begin{align}
    \lim_{k_1\to 0}\braket{\{\pi_1^g,\zeta_2^g,\zeta_3^g\}}'&=-\frac{(3-k_3\del_{k_3})\Im\psi_{2}(k_3)}{4(\Re\psi_{2}(k_3))^2}\\
    \lim_{k_1\to 0}\braket{\{\pi_1^g,\pi_2^g,\zeta_3^g\}}'&=\frac{\Re[\psi_{2}(k_3)(3-k_3\del_{k_3})\psi_{2}(k_3)^*]}{4(\Re\psi_{2}(k_3))^2}\\
    \lim_{k_1\to 0}\braket{\pi_1^g\pi_2^g\pi_3^g}'&=-\frac{\Im[(\psi_{2}(k_3))^2(3-k_3\del_{k_3})\psi_{2}(k_3)^*]}{4(\Re\psi_{2}(k_3))^2}.
\end{align}
You can see that Eqs.~\eqref{eq:bis_zzp}--\eqref{eq:bis_ppp} reduce to the above expressions by taking squeezed limit and removing long mode contributions $\psi_{2}({k}_\text{L})$. In the end, we explicitly confirmed that the local observables do not contain correlations between long and short modes of conjugate momenta as well as curvature perturbations. The IR regularity of the bubble loop at the boundary in the main text is implied by the above calculations.

\section{Cancellation in wavefunction for tensor modes}\label{sec:loe_tensor}
Here, we extend the analysis in Section \ref{sec:wave_ir} to wavefunction including tensor modes.
The wavefunction in geodesic coordinate is obtained through
\begin{align}
    \Psi^g[\zeta^g,\gamma^g]& \equiv\Psi[\zeta(\zeta^g),\gamma(\gamma^g)]\notag\\
    &=\mc{N}\exp\Bigg[-\frac{1}{2}\int_{\mf{k}_1,\mf{k}_2}(2\pi)^3\delta^{(3)}(\mf{k}_1+\mf{k}_2)\qty(\zeta_1^g\zeta_2^g\psi_2^g+\gamma^g_{1,ij}\gamma_{2,kl}^g\psi_2^{g,ijkl})\notag\\
    &\qquad-\frac{1}{6}\int_{\mf{k}_1,\mf{k}_2,\mf{k}_3}(2\pi)^3\delta^{(3)}\notag(\mf{k}_1+\mf{k}_2+\mf{k}_3)\notag\\
    &\qquad\quad {\times}\qty(\zeta_1^g\zeta_2^g\zeta_3^g\psi_3^g{+}3\zeta_2^g\zeta_2^g\gamma_{3,ij}^g\psi_3^{g,ij}{+} 3\zeta_1^g\gamma_{2,ij}^g\gamma_{3,kl}^g\psi_3^{g,ijkl}{+}\gamma_{1,ij}^g\gamma_{2,kl}^g\gamma^g_{3,nm}\psi_3^{g,ijklnm}){+}\cdots\Bigg],
\end{align}
where we used semi-classical approximation.\footnote{The loop contribution would contributes quadratic mixing $\psi_2^{ij}\zeta \gamma_{ij}$. Although we expect that the one-loop effects are sub-leading in IR due to necessity of derivative interactions $\del_i \zeta \del_j \zeta \gamma_{ij}$, $\zeta\dot{\gamma}_{ij}\dot{\gamma}_{ij}$, etc., we left the consistency relations and the analysis in local observables beyond semi-classical approximation of the wavefunction to future projects.}
In a geodesic coordinate system,
\begin{align}
    \label{eq:loe_zeta}
    \zeta^g_\mf{k}&\simeq\zeta_{\mf{k}}+\zeta_\text{L}(0)\qty(3+k\frac{\del}{\del k})\zeta_{\mf{k}}+\gamma_{\text{L},ab}(0)k^ak^b\frac{\del}{\del k^2}\zeta_{\mf{k}}\\
    \gamma^g_{\mf{k},ij}&\simeq \gamma_{\mf{k},ij}+\zeta_\text{L}(0)\qty(3+k\frac{\del}{\del k})\gamma_{\mf{k},ij}+\gamma_{\text{L},ab}(0)k^ak^b\frac{\del}{\del k^2}\gamma_{\mf{k},ij}.
    \label{eq:loe_gamma}
\end{align}
is satisfied. This basis transformation of the expansion leads to
\begin{align}
    -\frac{1}{2}\int_{\mf{k}_1,\mf{k}_2}&(2\pi)^3\delta^{(3)}(\mf{k}_1+\mf{k}_2)\zeta_1^g\zeta_2^g\psi_2^g\notag\\
    =&-\frac{1}{2}\int_{\mf{k}_1,\mf{k}_2}(2\pi)^3\delta^{(3)}(\mf{k}_1+\mf{k}_2)\zeta_1\zeta_2\psi_2^g\notag\\
    &-\frac{1}{2}\int_{\mf{k}_1,\mf{k}_2,\mf{k}_3,k_1\le k_\text{L}}(2\pi)^3\delta^{(3)}(\mf{k}_1+\mf{k}_2+\mf{k}_3)\zeta_1\zeta_2\zeta_3\qty(3-k_3\frac{\del}{\del k_3})\psi_2^g(k_2)\notag\\
    &+\frac{1}{2}\int_{\mf{k}_1,\mf{k}_2,\mf{k}_3,k_1\le k_\text{L}}(2\pi)^3\delta^{(3)}(\mf{k}_1+\mf{k}_2+\mf{k}_3)\gamma_{1,ij}\zeta_2\zeta_3k^i_2k^j_2\frac{\del}{\del k_2^2}\psi_2^g(k_2)\notag\\
    &+\order{\psi_4}
\end{align}
for second order of scalar part, and
\begin{align}
    -\frac{1}{2}\int_{\mf{k}_1,\mf{k}_2}&(2\pi)^3\delta^{(3)}(\mf{k}_1+\mf{k}_2)\gamma_{1,ij}^g\gamma_{2,kl}^g\psi_2^{g,ijkl}\notag\\
    =&-\frac{1}{2}\int_{\mf{k}_1,\mf{k}_2}(2\pi)^3\delta^{(3)}(\mf{k}_1+\mf{k}_2)\gamma_{1,ij}\gamma_{2,kl}\psi_2^{g,ijkl}\notag\\
    &-\frac{1}{2}\int_{\mf{k}_1,\mf{k}_2,\mf{k}_3,k_1\le k_\text{L}}(2\pi)^3\delta^{(3)}(\mf{k}_1+\mf{k}_2+\mf{k}_3)\zeta_1\gamma_{2,kl}\gamma_{3,mn}\qty(3-k_3\frac{\del}{\del k_3})\psi_2^{g,klmn}(k_2)\notag\\
    &+\frac{1}{2}\int_{\mf{k}_1,\mf{k}_2,\mf{k}_3,k_1\le k_\text{L}}(2\pi)^3\delta^{(3)}(\mf{k}_1+\mf{k}_2+\mf{k}_3)\gamma_{1,ij}\gamma_{2,kl}\gamma_{3,mn}k^i_2k^j_2\frac{\del}{\del k_2^2}\psi_2^{g,klmn}(k_2)\notag\\
    &+\order{\psi_4}
\end{align}
for second order of tensor part.
Thus, $\psi_2$ is unchanged, and relation between $\psi_3$ and $\psi_3^g$ is
\begin{align}
    \lim_{k_1\to 0}\psi_3^g&=\lim_{k_1\to 0}\psi_3-\qty(3-k_2\frac{\del}{\del k_2})\psi_2(k_2)\\
    \lim_{k_1\to 0}\psi_3^{g,ij}&=\lim_{k_1\to 0}\psi_3^{ij}+k^i_2k^j_2\frac{\del}{\del k_2^2}\psi_2(k_2)\\
    \lim_{k_1\to 0}\psi_3^{g,ijkl}&=\lim_{k_1\to 0}\psi_3^{ijkl}-\qty(3-k_2\frac{\del}{\del k_2})\psi_2^{ijkl}(k_2)\\
    \lim_{k_1\to 0}\psi_3^{g,ijklmn}&=\lim_{k_1\to 0}\psi_3^{ijklmn}+k^i_2k^j_2\frac{\del}{\del k_2^2}\psi_2^{klmn}(k_2).
\end{align}
It exactly cancels with Ward identity shown in \cite{Pimentel:2013gza},
\begin{align}
    \lim_{k_1\to 0}\psi_3&=\qty(3-k_2\frac{\del}{\del k_{2}})\psi_2(k_2)\\
    \lim_{k_1\to 0}\psi_3^{ij}&=-k_2^ik_2^j\frac{\del}{\del k_{2}^2}\psi_2(k_2)\\
    \lim_{k_1\to 0}\psi_3^{ijkl}&=\qty(3-k_2\frac{\del}{\del k_{2}})\psi_2^{ijkl}(k_2)\\
    \lim_{k_1\to 0}\psi_3^{ijklmn}&=-k_2^ik_2^j\frac{\del}{\del k_{2}^2}\psi_2^{klmn}(k_2).
\end{align}
Thus, the cancellation happens for whole perturbations in canonical single field inflation.

\end{fmffile}
\end{document}

%% file: setting/setting.tex
\pdfoutput=1
\usepackage{pifont,cancel,comment}
\usepackage[utf8]{inputenc}
\usepackage{here}
\usepackage{subfigure}
\usepackage{subfiles}
\usepackage{graphicx,xcolor}
\usepackage{amsmath,amssymb,amsbsy,amstext,amsthm,simplewick,amsfonts}
\usepackage{slashed}
\usepackage{lmodern}
\usepackage{bm}
\usepackage{caption}
\usepackage{subcaption}
\usepackage{physics}
\usepackage{braket}
\usepackage[version=4]{mhchem}
\usepackage{wasysym}
\usepackage{upgreek}
\usepackage{textgreek}

\usepackage{cite}

\usepackage{hyperref}
\hypersetup{colorlinks=true,linkcolor=bluecyan,citecolor=red3,urlcolor=green3,pdfencoding=auto,linktocpage}

\usepackage[framemethod=default]{mdframed}
\newmdenv[backgroundcolor=gray!15,%
skipabove=5pt,%
skipbelow=5pt,%
leftmargin=2pt,%
rightmargin=2pt,%
innertopmargin=-6pt,%
innerbottommargin=5pt,%
innerleftmargin=5pt,%
innerrightmargin=5pt,%
splittopskip=0pt,%
splitbottomskip=0pt,%
linewidth=0pt,%
nobreak=true]%
{keyeqn}

\newmdenv[backgroundcolor=gray!15,%
skipabove=5pt,%
skipbelow=5pt,%
leftmargin=2pt,%
rightmargin=2pt,%
innertopmargin=-2pt,%
innerbottommargin=5pt,%
innerleftmargin=5pt,%
innerrightmargin=5pt,%
splittopskip=0pt,%
splitbottomskip=0pt,%
linewidth=0pt,%
nobreak=true]%
{keythrm}

\definecolor{lightgreen}{cmyk}{0.2, 0, 0.2, 0.2}
\definecolor{lightgray}{cmyk}{0.1,0.2,0,0.1}
\definecolor{lightgray2}{cmyk}{0.1,0.1,0,0.1}
\definecolor{greyish2}{rgb}{.96,.96,.96}
\definecolor{bluecyan}{RGB}{0, 100, 200}
\definecolor{blue3}{RGB}{31,119,180}
\definecolor{red3}{RGB}{214,39,40}
\definecolor{orange3}{RGB}{255,127,14}
\definecolor{green3}{RGB}{44,160,44}
\definecolor{red2}{RGB}{255,0,0}
\definecolor{green2}{RGB}{0,170,0}
\definecolor{blue2}{RGB}{0,128,255}
\definecolor{magenta2}{RGB}{191,64,191}
\definecolor{purple2}{RGB}{112,48,160}
\definecolor{orange2}{RGB}{255,192,0}
\definecolor{blue2}{RGB}{117,223,230}

\newcommand{\email}[1]{\footnote{\href{mailto:#1}{\nolinkurl{#1}}}}

\renewcommand{\eqref}[1]{(\ref{#1})}

\numberwithin{equation}{section}

\allowdisplaybreaks[1]
\newcommand{\mpl}{M_\mathrm{pl}}

\renewcommand{\d}{\text{d}}

\newcommand{\mc}[1]{\mathcal{#1}}
\newcommand{\mr}[1]{\mathrm{#1}}

\newcommand{\mf}[1]{\mathbf{#1}}

\newcommand{\del}{\partial}

\newcommand{\pdif}[1]{\dfrac{\partial}{\partial #1}}

\newcommand{\vdif}[1]{\dfrac{\delta}{\delta #1}}

\newcommand{\vdiff}[2]{\dfrac{\delta #1}{\delta #2}}

\makeatletter

\@addtoreset{equation}{section}

\newcommand{\lambdabar}{{\mathchoice
  {\smash@bar\textfont\displaystyle{0.25}{1.2}\lambda}
  {\smash@bar\textfont\textstyle{0.25}{1.2}\lambda}
  {\smash@bar\scriptfont\scriptstyle{0.25}{1.2}\lambda}
  {\smash@bar\scriptscriptfont\scriptscriptstyle{0.25}{1.2}\lambda}
}}
\newcommand{\smash@bar}[4]{
  \smash{\rlap{\raisebox{-#3\fontdimen5#10}{$\m@th#2\mkern#4mu\mathchar'26$}}}%
}

\newcommand{\figcaption}[1]{\def\@captype{figure}\caption{#1}}
\newcommand{\tblcaption}[1]{\def\@captype{table}\caption{#1}}

\makeatother